\documentclass[a4paper,11pt]{article}
\pdfoutput=1 

\usepackage{jheppub} 
\usepackage[utf8]{inputenc}
\usepackage{setspace}
\usepackage{subfigure}
\usepackage{hyperref}
\definecolor{refkey}{gray}{0.45}
\definecolor{labelkey}{RGB}{155,48,48}

\usepackage{mathtools}
\usepackage{physics}
\usepackage{simplewick}
\usepackage[obeyFinal]{todonotes}
\usepackage[normalem]{ulem}

\def\beq{\begin{eqnarray}}\def\eeq{\end{eqnarray}}
\def\be{\begin{equation}}\def\ee{\end{equation}}

\def\mes[#1]{d^{3}{#1}}

\newcommand{\half}{\frac{1}{2}}

\def\order{\ensuremath{\mathcal{O}}}

\reversemarginpar

\definecolor{UI_blue}{RGB}{32, 64, 151}
\definecolor{UI_red}{RGB}{187, 62, 24}
\definecolor{UI_blue2}{RGB}{0, 84, 147}
\definecolor{UI_red2}{RGB}{159, 32, 66}
\definecolor{UI_gray}{RGB}{169, 169, 169}
\definecolor{UI_sepia}{RGB}{112, 66, 20}
\definecolor{UI_bittersweet}{RGB}{254, 111, 94}
\definecolor{UI_emerald}{RGB}{80, 200, 120}
\definecolor{UI_olivegreen}{RGB}{181, 179, 92}
\definecolor{UI_cadetblue}{RGB}{95, 158, 160}
\definecolor{UI_fuchsia}{RGB}{255, 0, 255}
\definecolor{UI_midnightblue}{RGB}{25, 25, 112}
\definecolor{UI_royalblue}{RGB}{0,35, 102}
\definecolor{UI_periwinkle}{RGB}{204, 204, 255}
\definecolor{UI_redorange}{RGB}{255, 83, 73}
\definecolor{UI_brickred}{RGB}{203,65,84}	
\definecolor{UI_forestgreen}{RGB}{34, 139, 34}
\definecolor{UI_tan}{RGB}{210,180,140}	
\definecolor{UI_burlywood}{RGB}{222,184,135}
\definecolor{UI_burlywood}{RGB}{192,64,0}
\definecolor{UI_darkorchid}{RGB}{153,50,204}

\newenvironment{system}%
  {\left\lbrace\begin{array}{@{}l@{}}}%
  {\end{array}\right.}

\usepackage{footmisc}
\usepackage{tikz}
\usetikzlibrary{arrows.meta}
\usepackage{titlesec}
\usepackage{verbatim}
\usepackage{fancyhdr}
\usepackage{enumerate}
\hypersetup{colorlinks=true,citecolor=blue,urlcolor=black}
\usepackage{slashed, enumitem}
\usepackage{array}
\newcolumntype{P}[1]{>{\centering\arraybackslash}p{#1}}
\usepackage{anyfontsize}
\usepackage{tikz}
\begin{document}

\begin{flushright}
	\hfill{YITP-24-118}
\end{flushright}

	\author[a, b]{Norihiro Iizuka,}
	\author[c]{Arkaprava Mukherjee,}
	\author[b]{Sunil Kumar Sake,}
	\author[b]{Nicol\`o Zenoni}

	\affiliation[a]{\it Department of Physics, National Tsing Hua University, Hsinchu 30013, Taiwan}
	\affiliation[b]{\it Yukawa Institute for Theoretical Physics, Kyoto University, Kyoto 606-8502, Japan}
	\affiliation[c]{\it Department of Physics, The Ohio State University, Columbus, OH 43210, USA}

	\emailAdd{iizuka@phys.nthu.edu.tw}
	\emailAdd{mukherjee.210@osu.edu}
	\emailAdd{sunilsake1@gmail.com}
	\emailAdd{nicolo@yukawa.kyoto-u.ac.jp}

	\vspace{1cm}

	\abstract{In quantum field theories that admit gravity dual, specific inequalities involving entanglement entropy between arbitrary disjoint spatial regions hold. An example is the  negativity of tripartite information.  Inspired by this, we investigate the analogous entropy inequalities in Sachdev-Ye-Kitaev (SYK) and sparse SYK models, which involve the entanglement among different flavors of Majorana fermions rather than spatial entanglement. Sparse SYK models are models where some of the SYK couplings are set to zero. Since these models have been argued to admit gravity duals up to a certain sparseness, it is interesting to see whether the multipartite entanglement structure changes in a sparseness-dependent manner. In the parameter space explored by our numerical analysis, which we performed upto five parties, we find that all entropy inequalities are satisfied for any temperature and degree of sparseness for an arbitrary choice of flavor subregions. In addition, if we plot the multipartite entanglement entropy in terms of purity, the only significant effect of sparseness is to change the range of purity. 
Thus, we conclude that multipartite information is almost unaffected by sparseness. 
As a counterexample, we also show that in a vector model of $N$-flavored Majorana fermions which contains no random variables, choices of subregions exist for which the entropy inequalities are violated. 
}

\title{Multipartite information in sparse SYK models}

	\maketitle
	
	\flushbottom
	
	\vskip 10pt
	
\section{Introduction}

The Sachdev-Ye-Kitaev (SYK) model \cite{Sachdev:1992fk,Kitaev:talk} contains all-to-all interactions of $N$-flavored Majorana fermions with random couplings. 
In the large-$N$ limit, $N\rightarrow\infty$, the SYK model at low energies captures the key properties of two-dimensional Jackiw-Teitelboim (JT) gravity \cite{Jackiw:1984je,Teitelboim:1983ux}, and thus represents a toy model of holography \cite{Kitaev:talk,Maldacena:2016hyu,Polchinski:2016xgd}. 
In the large-$N$ limit, when four-fermion interactions are considered, the SYK Hamiltonian is comprised of $\mathcal{O}(N^{4})$ terms and thus contains $\mathcal{O}(N^{4})$ random variables. 
A computationally simpler version of this model, referred to as the sparse SYK model, has been introduced in \cite{Xu:2020shn}. 
In the sparse SYK model, terms in the full Hamiltonian are switched off with probability $1-p$, so that $\mathcal{O} (p N^4)$ random variables are retained.\footnote{In our terminology, the full SYK model is the sparse SYK model with $p=1$.}
It has been argued that if $p > \mathcal{O}(N^{-3})$, or equivalently if $\mathcal{O}(N^a)$ terms are retained with $a > 1$,  the model still admits a gravity dual \cite{Xu:2020shn,Garcia-Garcia:2020cdo,Anegawa:2023vxq,Orman:2024mpw}. 
In particular, if $k N$ terms are left in the Hamiltonian, based on a path integral approach and a numerical analysis, in \cite{Xu:2020shn} it is argued that the sparse SYK manifests a maximally chaotic gravitational sector at low temperatures for $k > k_{\rm min}$, where $k_{\rm min} > 1/q$. For details of related analyses trying to estimate $k_{\rm min}$, see \cite{Xu:2020shn,Anegawa:2023vxq,Orman:2024mpw}. 
However, despite of these studies, it is not yet completely clear whether the sparse SYK model admits a gravity dual when the Hamiltonian contains only $\order{\left(N\right)}$ terms. 

Besides these results, it would be interesting to seek new smoking guns for the existence of a gravitational dual of the sparse SYK model. In this paper, we study the multipartite information in various sparse SYK models and try to understand the role of sparseness in them. 
To explain our motivation, let us briefly shift our focus to quantum field theories,
where entanglement entropy has been found to play a pivotal role as a probe of holography.
For a given state $\rho$ defined on a QFT time-slice,
the state of a spatial region $A$ is described by the reduced density matrix $\rho_A = \tr_{\bar{A}} \rho$,
with $\bar{A}$ the complementary region.
The entanglement entropy of the region $A$,
depending on both the state and the choice of the region itself,
is defined as the von Neuman entropy of $\rho_A$, 
namely $S(A) = - \tr_A (\rho_A \log \rho_A)$. 
Given two spatial regions $A$ and $B$, 
their correlations can be measured by the so-called mutual information, defined as
\begin{align}
\label{eq:def-mutual-info}
I_2(A,B) = S(A) + S(B) - S(AB) \, ,
\end{align}
in which $AB$ is a shorthand for $A \cup B$.
In any quantum system, the mutual information is positive, $I_2(A,B) \geq 0$, 
and monotonic, $I_2(A, BC) \geq I_2(A,B)$.
Equivalently, the entanglement entropy is strongly subadditive,
i.e. $S(AB) + S(BC) \geq S(ABC) + S(B)$.
Given three spatial regions $A,B,C$, the notion of tripartite information can be defined as
\begin{align}
\label{eq:def-tripartite-info}
I_3(A,B,C) = S(A) + S(B) + S(C) - S(AB) - S(BC) - S(AC) + S(ABC) \, .
\end{align}
Contrary to the mutual information $I_2(A,B)$, the tripartite information $I_3(A,B,C)$ in QFTs can have different signs according to the choice of the regions $A,B,C$ \cite{Casini:2008wt}.

For QFTs with a gravitational dual, the entanglement entropy of a spatial region 
can be computed by the Ryu-Takayanagi (RT) surface, {\it i.e.,} the area of extremal surfaces in the higher-dimensional bulk spacetime \cite{Ryu:2006bv,Ryu:2006ef,Nishioka:2009un,Hubeny:2007xt}. 
Such a holographic notion of entanglement entropy has been proven to satisfy the strong subadditivity \cite{Headrick:2007km}. Moreover, contrary to what happens for general QFTs, the tripartite information obtained from the holographic entanglement entropy is always negative for any choice of spatial regions $A,B,C$, a fact which is known as monogamy of holographic mutual information \cite{Hayden:2011ag}.
One of the implications of this finding is that,
relaying on the holographic approach to determine the entanglement entropy,  
the negativity of $I_3(A,B,C)$ is necessary for a QFT to admit a gravity dual.

In light of this, the question raises whether there are further inequalities satisfied by the holographic entanglement entropy that can be regarded as necessary conditions for the existence of a gravity dual.
To this purpose, one may look at the multipartite information
\begin{align}
\label{eq:def-multipartite-info}
I_\mathcal{N} (A_1, \dots, A_{\mathcal{N}}) = \sum_{s} (-1)^{|s|+1} S(s) \, ,
\end{align}
where $s$ denotes any possible subset of $\{ A_1, \dots, A_{\mathcal{N}} \}$
and $|s|$ the corresponding cardinality.
Note that for $\mathcal{N} =2,3$,
this definition corresponds to the mutual information in eq.~\eqref{eq:def-mutual-info}
and to the tripartite information in eq.~\eqref{eq:def-tripartite-info}, respectively.
As we have seen, for holographic QFTs $I_2(A_1,A_2) \geq 0$ and $I_3(A_1,A_2,A_3) \leq 0$.
However, if $\mathcal{N} \geq 4$ the multipartite information can have either signs according to the choice of spatial regions. 
Consequently, in general, the holographic entropy inequalities for more than three spatial regions
take a more complicated form.

A systematic study of the holographic entropy inequalities has been initiated in \cite{Bao:2015bfa},
and continued in \cite{Hubeny:2018trv,Hubeny:2018ijt,He:2019ttu,HernandezCuenca:2019wgh,Czech:2019lps,Avis:2021xnz,Czech:2021rxe,Czech:2022fzb,Czech:2023xed,Hernandez-Cuenca:2023iqh,Grado-White:2024gtx,Czech:2024rco,Bao:2024obe,Bao:2024vmy}.
In \cite{Bao:2015bfa} it has been shown that for $\mathcal{N} \leq 4$ there are no additional inequalities other than
the strong subadditivity of the entropy and the monogamy of the mutual information.
For higher number of spatial regions, the situation is more involved.
For example, for $\mathcal{N} = 5$ eight classes of inequalities have been found \cite{Bao:2015bfa} and have been shown to be a complete set \cite{HernandezCuenca:2019wgh}, 
while for $\mathcal{N}= 6$ additional classes of inequalities have been pointed out \cite{Hernandez-Cuenca:2023iqh}. 
Such inequalities take a simpler form when written in terms of the multipartite information quantities in eq.~\eqref{eq:def-multipartite-info} \cite{Hubeny:2018ijt,He:2019ttu,Hernandez-Cuenca:2023iqh}.

The above analysis involves the notion of entanglement entropy for spatial regions in QFTs.
On the other hand, the (sparse) SYK model is a $(0+1)-$dimensional theory 
that has no 'space'. Therefore, the only notion of entanglement entropy that can be applied in this model is the entanglement in the flavor space. 
Equivalently, for a given state $\rho$ of the quantum system, 
a reduced density matrix $\rho_{A}$ for a subsystem in the Hilbert space 
can be obtained by tracing over degrees of freedom outside the selected subsystem in the flavor space. 
Again, the entanglement entropy is nothing but the von Neumann entropy $S(A) =- \tr_A(\rho_{A} \log \rho_{A})$.
We stress that this definition of entanglement entropy is physically different from
the entanglement entropy for spatial regions.
In particular, even though the SYK model is known to admit a gravity dual,
no gravitational counterpart for the Hilbert space entanglement entropy is yet known.
This is simply because SYK model is a quantum mechanical model. 

In this paper, we investigate the analog of the multipartite information defined in eq.~\eqref{eq:def-multipartite-info}
for the (sparse) SYK models, and we check whether the holographic entropy inequalities are satisfied by these quantities. 
For comparison, we also consider two different types of non-random Hamiltonian models with four fermi-interactions.  Following are the Hamiltonians of various models we consider in this paper:
\begin{align}
	H_{\text{SYK}}&= \sum_{1 \leq a<b<c<d \leq N}  \mathcal{P}_{abcd} \, J_{abcd} \, \psi_a \psi_b \psi_c \psi_d \, ,\\
	 H_{V1} &= \left( \sum_{a=1}^{N/2} \psi_{2a-1} \psi_{2a} \right)^2 \,, \\
	 H_{V2} & = \left( \sum_{a=1}^{N/2} \psi_{a} \psi_{N/2+a} \right)^2 \, ,
	 \label{hamvar}
\end{align}
where $J_{abcd}$ are random couplings drawn from the Gaussian distribution in eq.~\eqref{pdfj} and $\psi_a$ are the Majorana fermions satisfying the Clifford algebra \eqref{eq:clifford}. The parameter $\mathcal{P}_{abcd}$ can take value $1$ with probability $p$ or value $0$ with probability $1-p$, with $0 < p \leq 1$.

In all the three models,  we define the $N/2$ spin degrees of freedom as
\begin{align}
\label{eq:spin-base}
\psi_{2a} = \frac{c_a + \bar{c}_a}{\sqrt{2}} \, ,
\qquad 
\psi_{2a-1} = \frac{i (c_a - \bar{c}_a)}{\sqrt{2}}   \, ,
\qquad
a=1, \dots, N/2 \, ,
\end{align}
and we divide the whole system into $M$ subregions, where the subregions $A_j$ contains $n_j$ spins, {\it i.e.,} $A_j$ is defined as 
\begin{align}
\label{eq:def-Aj}
A_j = \{ n_1 + \dots + n_{j-1} +1, \dots, n_1 + \dots +n_{j-1} + n_{j} \} \, ,
\qquad
\sum_{j=1}^{M} n_j = N/2 \, ,
\end{align}
 The main findings can be summarized as follows.
\begin{itemize}
	\item  For the full SYK model, all the entropy inequalities up to $\mathcal{N} =5$, that are valid in higher dimensional holographic QFTs, hold. For $\mathcal{N}= 6$, there is a large list of entropy inequalities, which we partially check to also hold.   
	\item For the sparse SYK model, even for the case of extreme sparseness, when the effective number of terms in the Hamiltonian is less than or equal to $\mathcal{O}(N)$, in which case the holographic dual is not expected to exist, the tripartite information $I_3$ is still negative, suggesting that its negativity is not a sufficient condition to test the holographic nature of the sparse SYK model. 
	\item For the non-random models $H_{V1},H_{V2}$, we find that they are different from the perspective of multipartite information.  The value of the tripartite information is positive in the model $H_{V1}$ and negative in the model $H_{V2}$. 
Note that $H_{V1}$ and $H_{V2}$ are the same, up to a relabeling of the Majorana fermions. 
Indeed, by starting with the Hamiltonian $H_{V1}$ and changing the way subsystems $A_j$ are defined, compared to  \eqref{eq:def-Aj}, it is possible to obtain the Hamiltonian $H_{V2}$ with the choice \eqref{eq:def-Aj}.
Therefore, we conclude that in the non-random model, the sign of $I_3$ is affected by the choice of subregions.
	\item Moreover, the five-party inequalities are violated by $H_{V1}$ and satisfied by $H_{V2}$.  
	Interestingly, when $I_3$ is positive, the other holographic inequalities are violated as well. 
	Whether this phenomenon is more general is beyond the scope of this work. 
\end{itemize}

The rest of the paper is organized as follows. In section \ref{sec:setup} we review the sparse SYK model and the nature of flavor space entanglement entropy in this model. In section \ref{sec:tripartite-info} we elaborate on the nature of tripartite information in our models. The nature of higher partite information is discussed in section \ref{sec:multipartite-info}. In section \ref{sec:EE-inequalities} we check the various entropy inequalities for $\mathcal{N} =5$ parties. We end in section \ref{sec:conclusions} with conclusions and some open questions. Details are deferred to the appendices.


\section{Setup}
\label{sec:setup}

	Let us begin with a brief review of the sparse SYK model with a view towards some of the properties that will be important to understand the nature of flavor space entanglement in this model.

\subsection{Sparse SYK}

We consider the following Hamiltonian
 \begin{align}
 \label{eq:H-syk}
 H = \sum_{1 \leq a<b<c<d \leq N}  \mathcal{P}_{abcd} \, J_{abcd} \, \psi_a \psi_b \psi_c \psi_d \, ,
 \end{align}
where $\psi_a$ are $N$-flavored Majorana fermions satisfying the Clifford algebra\footnote{A convenient basis of matrices satisfying the Clifford algebra is formed by tensorial products of $n = N/2$ Pauli matrices.
In particular, we can take
\begin{align}
\psi_{2a-1} &= \frac{1}{\sqrt{2}} \, \overset{a-1}{\underset{l=1}\otimes} \sigma_z^l \otimes \sigma_x^a \overset{n}{\underset{l=a+1}\otimes} \mathbb{I}^l \, , \label{eq:psi-odd} \\
\psi_{2a} &=  \frac{1}{\sqrt{2}} \, \overset{a-1}{\underset{l=1}\otimes} \sigma_z^l \otimes \sigma_y^a \overset{n}{\underset{l=a+1}\otimes} \mathbb{I}^l \, , \label{eq:psi-even}
\end{align}
with $a = 1, \dots, n$.}
\begin{align}
\label{eq:clifford}
\{\psi_a,\psi_b\}=\delta_{ab} \, , 
\qquad
a,b = 1, \dots, N.
\end{align}
The variable $J_{abcd}$ is taken from a Gaussian distribution with probability
\begin{align}
p(J_{abcd}) = \frac{1}{\sqrt{2 \pi J^2}} e^{- \frac{J_{abcd}^2}{2 J^2}} \, ,\label{pdfj}
\end{align}
so that
\begin{align}
\langle J_{abcd} J_{a'b'c'd'} \rangle = \frac{J^2}{p} \delta_{aa'} \delta_{bb'} \delta_{cc'} \delta_{dd'} \, , \qquad J^2=\frac{6}{N^3}\label{jvar}
\end{align}
The parameter $\mathcal{P}_{abcd}$ is a boolean variable drawn from a binary distribution with probability
\begin{align}
\mbox{probability}(\mathcal{P}_{abcd}) =
\begin{cases}
	p \qquad &\text{if $\mathcal{P}_{abcd}=1$} \\
	1-p \qquad &\text{if $\mathcal{P}_{abcd}=0$} 
\end{cases}
\label{pbin}
\end{align}
with $0<p \leq 1$.
For $p=1$, we have the usual SYK model with interaction term $\left( \psi \right)^{4}$.
Instead, for $0<p <1$ some terms in the usual SYK Hamiltonian are switched off with probability $1-p$. 
Such a model is referred to as a sparse SYK model.
The average number of terms in Hamiltonians of the form \eqref{eq:H-syk} is
\begin{align}
\#{\rm terms} = p \binom{N}{4} \, .\label{efterm}
\end{align}
In the following, we specialize to $N = 2n$ Majorana fermions. 
Equivalently, we can introduce $n$ Dirac fermions $c_a$ by
\begin{align}
\label{eq:creator-annihilator}
\psi_{2a} = \frac{c_a + \bar{c}_a}{\sqrt{2}} \, ,
\qquad 
\psi_{2a-1} = \frac{i (c_a - \bar{c}_a)}{\sqrt{2}}   \, ,
\end{align}
which satisfy the anti-commutation relations
\begin{align}
\{ c_a , \bar{c}_b \} = \delta_{ab} \, ,
\qquad
\{ c_a , c_b \} = \{ \bar{c}_a , \bar{c}_b \} = 0 \, .\label{ccomms}
\end{align}
It is here convenient to consider the spin basis.
The operators $c_a, \bar{c}_a$ act on the $a$-th spin as annihilation and creation operators, respectively:
\begin{align}
c_a \ket{\uparrow}_a = \ket{\downarrow}_a \, , 
\qquad
\bar{c}_a \ket{\downarrow}_a = \ket{\uparrow}_a \, , 
\qquad
c_a \ket{\downarrow}_a = \bar{c}_a \ket{\uparrow}_a =0 \, .
\end{align}
In the following, we will also make use of the number operator
$N_a = \bar{c}_a c_a$, acting as 
$N_a \ket{\uparrow}_a = \ket{\uparrow}_a$ and $N_a \ket{\downarrow}_a = 0$.

\subsection{Degeneracy of the spectrum}
\label{subsec:degeneracy}

\begin{figure}[h!]
\begin{center}
\begin{tikzpicture}
 \draw (0,-0.2)--(0,2);
 \draw (-1.5,0.5)--(-0.5,0.5);
 \draw (0.5,0.5)--(1.5,0.5);
 \draw (-1.5,1.2)--(-0.5,1.2);
 \draw (0.5,1.2)--(1.5,1.2);
 \node[below]  at (-1,0.4) {$\ket{GS}_1$};
 \node[below]  at (1,0.4) {$\ket{GS}_2$};
 \node  at (-1,1.7) {$\dots$};
 \node  at (1,1.7) {$\dots$};
 \node[above] at (-1,2.1) {$+$};
 \node[above] at (1,2.1) {$-$};
 \node[below] at (0,-0.5) {\bf $N$ mod $4=2$};
\end{tikzpicture}
\qquad \quad
\begin{tikzpicture}
 \draw (0,-0.2)--(0,2);
 \draw (-1.5,0.5)--(-0.5,0.5);
 \draw (-3,0.5)--(-2,0.5);
 \draw (2,1.2)--(3,1.2);
 \draw (0.5,1.2)--(1.5,1.2);
 \node[below]  at (-2.5,0.4) {$\ket{GS}_1$};
 \node[below]  at (-1,0.4) {$\ket{GS}_2$};
 \node  at (-1.7,1.7) {$\dots$};
 \node  at (1.7,1.7) {$\dots$};
 \node[above] at (-1.7,2.1) {$+$};
 \node[above] at (1.7,2.1) {$-$};
 \node[below] at (0,-0.5) {\bf $N$ mod $8=4$};
\end{tikzpicture}
\qquad \quad
\begin{tikzpicture}
 \draw (0,-0.2)--(0,2);
 \draw (-1.5,0.5)--(-0.5,0.5);
 \draw (0.5,1.2)--(1.5,1.2);
 \node[below]  at (-1,0.4) {$\ket{GS}$};
 \node  at (-1,1.7) {$\dots$};
 \node  at (1,1.7) {$\dots$};
 \node[above] at (-1,2.1) {$+$};
 \node[above] at (1,2.1) {$-$};
 \node[below] at (0,-0.5) {\bf $N$ mod $8=0$};
\end{tikzpicture}
\end{center}
\caption{Schematic representation of the degeneracy and chirality of the energy eigenstates. Note that in the second and third case, the chirality of the ground state(s) depends on the $\mathcal{P}_{jlmn}, J_{jlmn}$ variables in the Hamiltonian.}
\label{fig:degeneracy}
\end{figure}
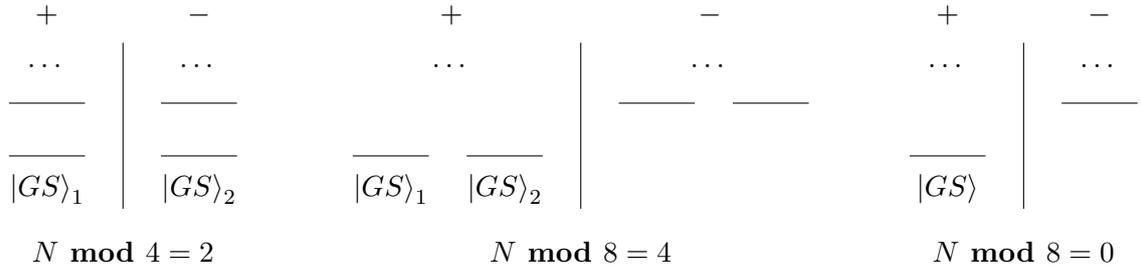

The degeneracy of the spectrum of the (sparse) SYK model is known, see for example \cite{Fidkowski:2010jmn,Fu:2016yrv,You:2016ldz,Cotler:2016fpe,Garcia-Garcia:2020cdo}.
Here we collect the salient results that will be useful in our discussion,
and we relegate the details to Appendix \ref{app:degeneracy}.
For the full SYK model, the degeneracy of the spectrum depends on the number of Majorana fermions $N$.
In particular, if $N$ mod $8 \neq 0$ a two-fold degeneracy is present, whereas if $N$ mod $8 =0$ the degeneracy is not guaranteed, see Fig.~\ref{fig:degeneracy}.
The $+$ and $-$ denotes different chirality sectors, where the chirality operator is 
\begin{align}
\gamma = (2 i)^n \prod_{a=1}^{2n} \psi_a
 = \prod_{a=1}^n ( 2 N_a -1) \, ,
 \end{align}
 whose spectrum is $\{ +1, -1 \}$ and $\left[ \gamma, H \right]=0$.
 
 With the increasing of sparseness, or equivalently the decreasing of $p$, less terms in the Hamiltonian \eqref{eq:H-syk} are retained. Therefore, the number of symmetries increases. Consequently, as discussed in \cite{Garcia-Garcia:2020cdo}, the degree of degeneracy of the spectrum can increase. For instance, in the extremely sparse case where the Hamiltonian contains just one term, there are just two energy levels, each of which has degeneracy $2^{N/2-1}$. 
In general, finding the degree of degeneracy for a given value of $p<1$  is a challenging task.

\subsection{Entanglement of flavors}
\label{subsec:entropy-GS}

We can define a notion of entanglement entropy in the flavor space between different flavors of Majorana fermions.
Let us first pair the $N$ Majorana fermions into $n=N/2$ Dirac fermions as in eq.~\eqref{eq:creator-annihilator}.  
Then, we refer to a subgroup of $n_A$ Dirac fermions as subregion $A$, and to the remaining $n_{\bar{A}} = n - n_A$ as subregion $\bar{A}$. 
Thus, if the full system of $n$ spins is in a state $\rho$, the reduced density matrix for $A$ can be obtained by tracing over the $n_{\bar{A}}$ subgroup:
\begin{align}
\rho_A = \tr_{\bar{A}} \rho \, .
\end{align}
The entanglement entropy $S_A$ between the two subgroups of fermions is given by the Von neumann entropy of the reduced density matrix $\rho_A$:
\begin{align}
S_A = - \tr_A (\rho_A \log \rho_A) \, .
\end{align}

\subsubsection{Ground state}
\begin{figure}[h]
\centering
\includegraphics[scale=0.7]{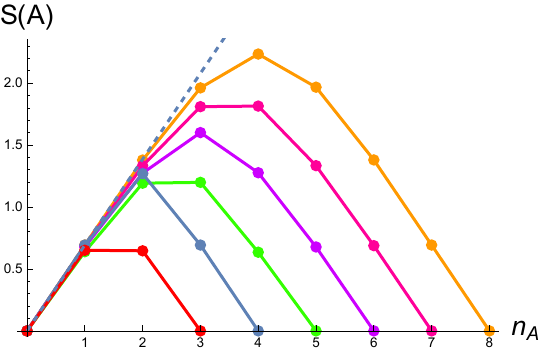} 
\caption{Entanglement entropy for the typical ground state $\rho_{GS} = \ketbra{\psi}$ in eq.~\eqref{eq:typical-GS} for the SYK model as a function of the subregion size $n_A$. Each curve is obtained by an ensemble average $S_A = \sum_{1 \leq i \leq N_{\rm ens}}S_{A,i}/N_{\rm ens}$ for fixed number of Majorana fermions $N=6,8,10,12,14,16$. For any case we have considered $N_{\rm ens}=100$. The dotted line represents $S_A = n_A \log2$. The plot for a single ground state $\rho_{GS} = \ketbra{GS_1}$ is the same. See also \cite{Fu:2016yrv}.}
\label{fig:entropy-GS}
\end{figure}

When $\rho = \ketbra{\psi}$ is a pure state, we have $S_{A \cup \bar{A}} =0$ and $S_A = S_{\bar{A}}$.
A plot of the entanglement entropy $S_A$ as a function of the subregion size $n_A$ is shown in Fig. \ref{fig:entropy-GS}. The curve is drawn for a typical ground state of SYK, obtained by a random linear combination of the various ground states in any instantiation, mathematically given by 
\begin{align}
\label{eq:typical-GS}
\ket{\psi} = \sum_{j=1}^g e^{i \phi_j} \omega_j \ket{GS_j} \, ,
\qquad
\sum_{j=1}^g \omega_j^2 =1 \, .
\end{align}
Here, $g$ is the degeneracy of the ground state (see Appendix \ref{app:degeneracy}), $\phi_j , \omega_j$ are real variables in the ranges $0 \leq \phi_j < 2 \pi$, 0 $\leq \omega_j \leq 1$, and the states $\ket{GS_j}$ are the set of degenerate orthonormal ground states in any instance of the theory. For each ensemble, the values of $\phi_j, \omega_j$ are randomly taken from a uniform distribution. 

As already observed in \cite{Fu:2016yrv}, the entanglement entropy follows the Page curve \cite{Page:1993df}.

\subsubsection{Thermal state}
\label{subsec:entropy-thermal}
Apart from the entanglement structure of the ground state, it is also interesting to understand the nature of entanglement {vis a vis} the temperature of the system. So, we now turn our attention to another well-known state, namely the thermal state of the system. 
For that purpose, we consider the full system of $N=2n$ Majorana fermions to be in the thermal state.
By exact diagonalization of the Hamiltonian \eqref{eq:H-syk}, we can obtain the $d= 2^n$ orthonormal eigenstates and corresponding energies: $H \ket{E_j} = E_j \ket{E_j}$, with $j= 1, \dots, d$.
Then, the thermal state can be explicitly built as
\begin{align}
\label{eq:thermal-state}
\rho_{\beta} = \frac{e^{- \beta H}}{Z_\beta} = \frac{1}{Z_\beta} \sum_{j=1}^d e^{- \beta E_j} \ketbra{E_j} \, ,
\qquad
Z_\beta = \sum_{j=1}^d e^{- \beta E_j} \, ,
\end{align}
where $\beta$ is the inverse temperature of the system and the state is normalized so that $\tr (\rho_\beta) =1$.
The thermal state is a mixed state, whose entanglement entropy coincides with the thermal entropy
\begin{align}
\label{eq:thermal-entropy}
S(\rho_\beta) = \beta \langle E \rangle_{\rm th} + \log Z_\beta \, , 
\end{align}
with $\langle E \rangle_{\rm th}$ the thermal average of the energy.
In the $\beta =0$ (infinite temperature) limit, the thermal state corresponds to the maximally mixed state
\begin{align}
\label{eq:rho-0}
\rho_0 = \frac{1}{d} \sum_{j=1}^d \ketbra{E_j} = \frac{1}{d} \, \mathbb{I}_d \, ,
\end{align}
with maximal entanglement entropy $S(\rho_0) = \log d = n \log2$.
In the $\beta \to +\infty$ (zero temperature) limit, just the contributions from the ground state(s) survive and so
\begin{align}
\label{eq:rho-infinity}
\rho_\infty = \frac{1}{g} \sum_{j=1}^g \ketbra{GS_j} \, .
\end{align}
In this case, the entanglement entropy acquires its minimal value $S(\rho_\infty) = \log g$.
Note that if $g=1$, $\rho_\infty$ is a pure state and the corresponding entropy vanishes.

\section{Tripartite information inequalities}
\label{sec:tripartite-info}

We are interested in the tripartite information introduced in eq.~\eqref{eq:def-tripartite-info}.
The tripartite information can be conveniently expressed in terms of the mutual information \eqref{eq:def-mutual-info} as
\begin{align}
\label{eq:I3}
I_3(A,B,C) = I_2(A,B) + I_2(A,C) - I_2(A,BC) \, .
\end{align}
For a general mixed state, the tripartite information $I_3$ can be either positive or negative.\footnote{As pointed out in \cite{Hayden:2011ag}, let us consider a system of three spins in two different states 
\begin{align}
\rho_1 = \frac{1}{2} \left( \ketbra{\uparrow \uparrow \uparrow} + \ketbra{\downarrow \downarrow \downarrow} \right)
\end{align} and 
\begin{align}
\rho_2 = \frac{1}{4} \left( \ketbra{\uparrow \uparrow \uparrow} + \ketbra{\uparrow \downarrow \downarrow} + \ketbra{\downarrow \uparrow \downarrow}+ \ketbra{\downarrow \downarrow \uparrow} \right) \, .
\end{align}
With the choice of $A,B,C$ as the first, second, and third spin respectively, 
we have $I_2(A,B) = I_2(A,C) = I_2(A,BC) = \log 2$ for the state $\rho_1$, and the redundancy of correlations leads to $I_3(A,B,C) > 0$.
For the state $\rho_2$ we have $I_2(A,B) = I_2(A,C) = 0$ and $I_2(A,BC) = \log 2$. The correlations between $A$ and $BC$ are not preserved when the latter is split into $B$ and $C$, leading to the monogamy relation $I_3(A,B,C) < 0$.
\label{footnote:I3}}
However, for a state in a QFT with a smooth gravity dual, the tripartite information $I_3$ is negative, a condition referred to as monogamy of mutual information \cite{Hayden:2011ag}.
Indeed, the negativity of the tripartite information is equivalent to
$I_2(A,B) + I_2(A,C) \leq I_2(A,BC)$,
a condition which is realized when the correlations between $A$ and $BC$ 
cannot be fully preserved when $BC$ is split into $B$ and $C$.
In the following, we want to investigate whether mutual information in the (sparse) SYK, 
for a splitting of the $N$ flavors into three subregions $A, B, C$, is monogamous.
We take the full system $ABC$ to be in the thermal state $\rho_\beta$ given in eq.~\eqref{eq:thermal-state} and the three subregions to contain the same even number of flavors $N/6$, or equivalently the same number of spins $n/3$. 
Namely, we choose
\begin{align}
\label{eq:ABC-tripartite}
A = \left\{ j \left| 1 \leq j \leq \frac{n}{3} \right. \right\} \, ,
\qquad
B = \left\{ j \left| \frac{n}{3} +1 \leq j \leq \frac{2n}{3} \right. \right\} \, ,
\qquad
C = \left\{ j \left| \frac{2n}{3}+1  \leq j \leq n \right. \right\} \, .
\end{align}

\begin{center}
\begin{tikzpicture}
 \draw[fill=blue!40!white] (0,0) circle [radius=0.5];
 \draw[-{Latex[length=2mm,width=2mm]}] (0,-0.6) to  (0,0.8);
 \node at (1,0) {$\dots$};
 \draw[fill=blue!40!white] (2,0) circle [radius=0.5];
 \draw[-{Latex[length=2mm,width=2mm]}] (2,-0.6) to  (2,0.8);
 \draw (-0.5,-0.8)--(-0.5,-0.9)--(2.5,-0.9)--(2.5,-0.8);
 \node[below] at (1,-0.9) {$A=n/3$ spins}; 
 \draw[fill=blue!40!white] (3.5,0) circle [radius=0.5];
 \draw[-{Latex[length=2mm,width=2mm]}] (3.5,0.6) to (3.5,-0.8);
 \node at (4.5,0) {$\dots$};
 \draw[fill=blue!40!white] (5.5,0) circle [radius=0.5];
 \draw[-{Latex[length=2mm,width=2mm]}] (5.5,-0.6) to  (5.5,0.8);
 \draw (3,-0.8)--(3,-0.9)--(6,-0.9)--(6,-0.8);
 \node[below] at (4.5,-0.9) {$B=n/3$ spins}; 
 \draw[fill=blue!40!white] (7,0) circle [radius=0.5];
 \draw[-{Latex[length=2mm,width=2mm]}] (7,-0.6) to  (7,0.8);
 \node at (8,0) {$\dots$};
 \draw[fill=blue!40!white] (9,0) circle [radius=0.5];
 \draw[-{Latex[length=2mm,width=2mm]}] (9,0.6) to (9,-0.8);
 \draw (6.5,-0.8)--(6.5,-0.9)--(9.5,-0.9)--(9.5,-0.8);
 \node[below] at (8,-0.9) {$C=n/3$ spins}; 
 \end{tikzpicture}
\end{center}

\subsection{Tripartite information for the sparse SYK}
\label{subsec:tripartite-info-syk}

Before analyzing the numerical results, let us discuss some limiting cases.
\\

{\bf Infinite temperature limit.}
When $\beta =0$, the full system $ABC$ is in the maximally mixed state $\rho_0$ in eq.~\eqref{eq:rho-0}, so $S_{ABC}= n \log2$.
When tracing out spin degrees of freedom, the resulting reduced density matrix is still the identity matrix with a lower dimensionality\footnote{Since the identity matrix is unaffected by a change of basis, $\rho_0$ is proportional to the identity both in the eigenvectors and in the spin basis. By working in the spin basis, tracing out a subsystem of $n_{\rm sub}$ spins results in a diagonal matrix in which each entry is the sum of $2^{n_{\rm sub}}$ distinct elements on the diagonal of the starting matrix $\rho_0$.} and a proper normalization factor. 
In the case at hand,
\begin{align}
\rho_{AB} = \rho_{BC} = \rho_{AC} = \frac{1}{2^{2n/3}} \, \mathbb{I}_{2^{2n/3}} 
\qquad
&\implies
\qquad
S_{AB} = S_{BC} = S_{AC} = \frac{2n}{3} \log 2 \, ,
\nonumber \\
\rho_A = \rho_B = \rho_C = \frac{1}{2^{n/3}} \, \mathbb{I}_{2^{n/3}} 
\qquad
&\implies
\qquad
S_A = S_B = S_C = \frac{n}{3} \log 2 \, .
\nonumber
\end{align}
Putting all together, the tripartite information for the maximally mixed state is
\begin{align}
I_3(\rho_0) = n \log 2 - 2 n \log 2 + n \log 2 
= 0 \, .
\end{align}
Note that this result holds for any general tripartition $n_A \neq n_B \neq n_C$.
\\

{\bf Large but finite temperature.}
In the limit $\beta \to 0^+$, both the thermal density matrix for the full system and its partial trace for the reduced system can be expressed as
\begin{align}
\label{eq:rho-small-beta}
\rho &\sim \frac{\mathbb{I} - \beta M_1 + \frac{\beta^2}{2} M_2}{\tr \left( \mathbb{I} - \beta M_1 + \frac{\beta^2}{2} M_2 \right)} 
= \frac{\mathbb{I} - \beta M_1 + \frac{\beta^2}{2} M_2}{d - \beta \tr (M_1) + \frac{\beta^2}{2} \tr (M_2)} \\
&\sim \frac{1}{d} \left( \mathbb{I} + \frac{\beta}{d} \left( \tr (M_1) \mathbb{I} - d \, M_1 \right) + \frac{\beta^2}{2 d^2} \left( (2 (\tr (M_1))^2 -d \tr M_2) \mathbb{I} - 2d \, \tr (M_1) M_1 + d^2 \, M_2  \right) \right) \, , \nonumber
\end{align}
where $d = \tr (\mathbb{I})$ is the dimension of the (sub)system on which $\rho$ is defined.
Note that $\rho$ is normalized up to $\mathcal{O}(\beta^2)$.
The entanglement entropy thus reads
\begin{align}
\label{eq:S-small-beta}
S(\rho) = - \tr (\rho \log \rho)
=  \log d + \frac{(\tr (M_1))^2 - d \tr (M_1^2)}{2 d^2} \beta^2 + \mathcal{O}(\beta^3) \, .
\end{align}
So, the leading correction to the entanglement entropy is quadratic in $\beta$ and all the information on such a correction is in the $\mathcal{O}(\beta)$ term of the (reduced) density matrix, namely in $M_1$ and its trace.
It can be easily checked that the leading correction to the entanglement entropy is negative.\footnote{Since the trace is basis-independent, let us consider the basis in which both $M_1$ and $M_1^2$ are diagonal:
$(M_1)_{ij} = m_i \, \delta_{ij}$ and $(M_1^2)_{ij} = m_i^2 \, \delta_{ij}$.
Then, we get
\begin{align}
\label{eq:S-neg}
(\tr (M_1))^2 - d \tr (M_1^2) = 
\left( \sum_{i=1}^d m_i \right)^2 - d \sum_{i=1}^d m_i^2
= \sum_{i=1}^d (1-d) m_i^2 + 2 \sum_{i<j} m_i m_j  
= - \sum_{i<j} (m_i - m_j)^2 \, ,
\end{align}
which proves our statement.}
\\
The tripartite information for large temperature turns out to be
\begin{align}
\label{eq:I3-small-beta}
I_3 (\rho_\beta) = \frac{\beta^2}{2 d_{ABC}^2} \left(
(\tr (H))^2 + \tr \left( 
- d_{ABC} \, H^2
- d_{A} \, H_{A}^2
- d_{B} \, H_{B}^2
\right. \right. \nonumber  \\
\left. \left.
- d_{C} \, H_{C}^2
+ d_{AB} \, H_{AB}^2
+ d_{BC} \, H_{BC}^2
+ d_{AC} \, H_{AC}^2
\right) \right)
+ \mathcal{O}(\beta^3) \, ,
\end{align}
where we have defined $H_A = \tr_{BC} (H)$ and so on.
Even if the leading correction to the entanglement entropy has always negative sign, the leading correction to $I_3$ at small $\beta$ for a general system is not sign definite.
For a detailed derivation of these results see Appendix \ref{app:I3-large-T}.
\\

{\bf Zero temperature limit.}
When $\beta \to +\infty$, the full system $ABC$ is in the state $\rho_\infty$ given in eq.~\eqref{eq:rho-infinity}.
If $N$ mod $8=0$ and $g=1$, the state is pure.
As we have discussed in subsection.~\ref{subsec:entropy-GS}, the entanglement entropy of a pure state vanishes and the entanglement entropy of a pure state restricted to a subregion equals the entanglement entropy of the complementary subregion.
Therefore,
\begin{align}
S_{ABC} =0 \, , 
\qquad
S_A = S_{BC} \, ,
\qquad
S_B = S_{AC} \, ,
\qquad
S_C = S_{AB} \, ,
\end{align}
implying that $I_3(\rho_{\infty,g=1}) =0$.
If $N$ mod $8 \neq 0$, then $g \geq 2$ \cite{Fidkowski:2010jmn,Fu:2016yrv,You:2016ldz,Cotler:2016fpe,Garcia-Garcia:2020cdo} and $\rho_{\infty,g}$ is not pure.
In this case, we have no simple argument to predict the sign of $I_3(\rho_{\infty,g})$.
\\

\begin{figure}[h]
\centering
\includegraphics[scale=0.7]{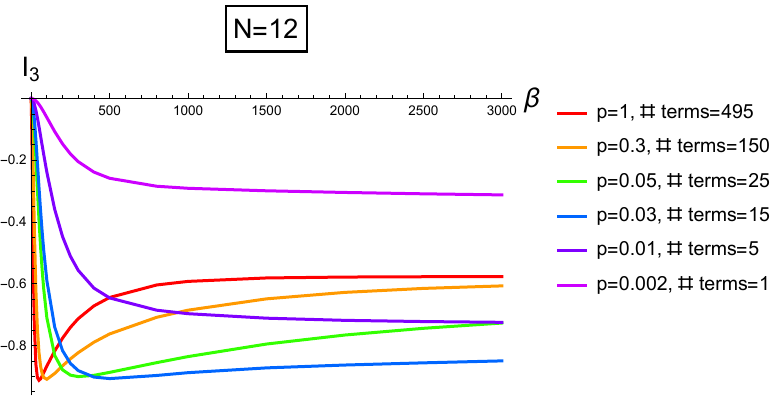} 
\\
\includegraphics[width=7cm,height=5cm]{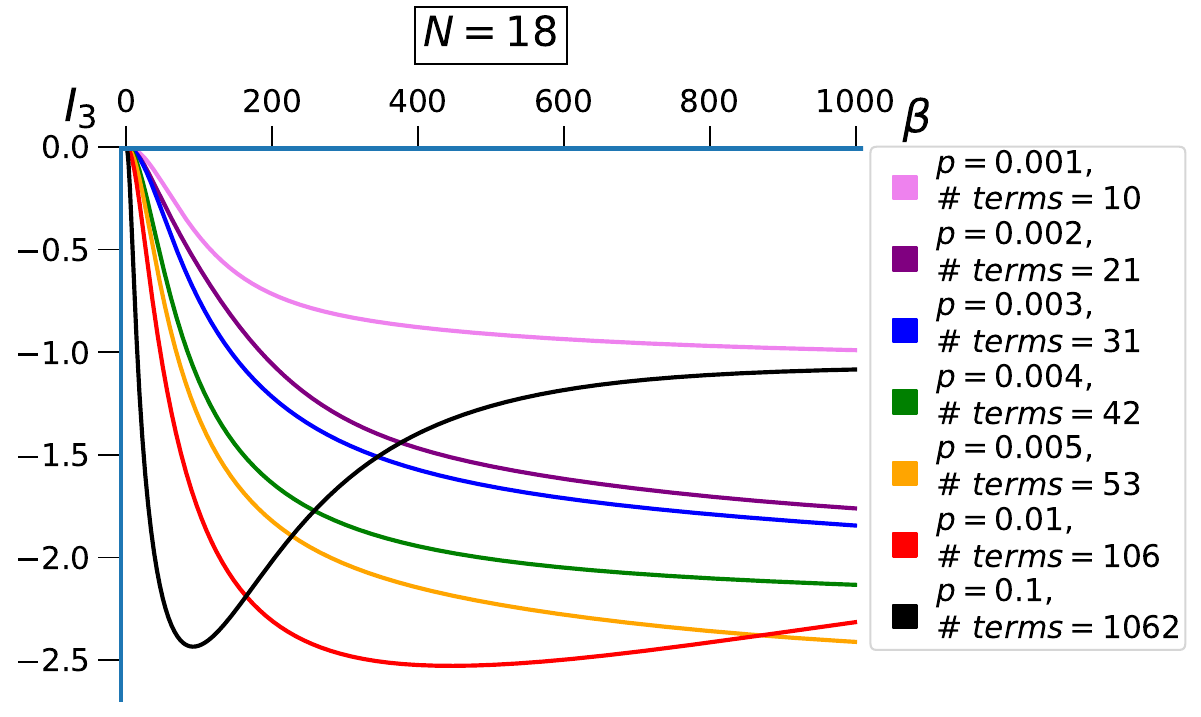} 
\qquad
\includegraphics[width=7cm,height=5cm]{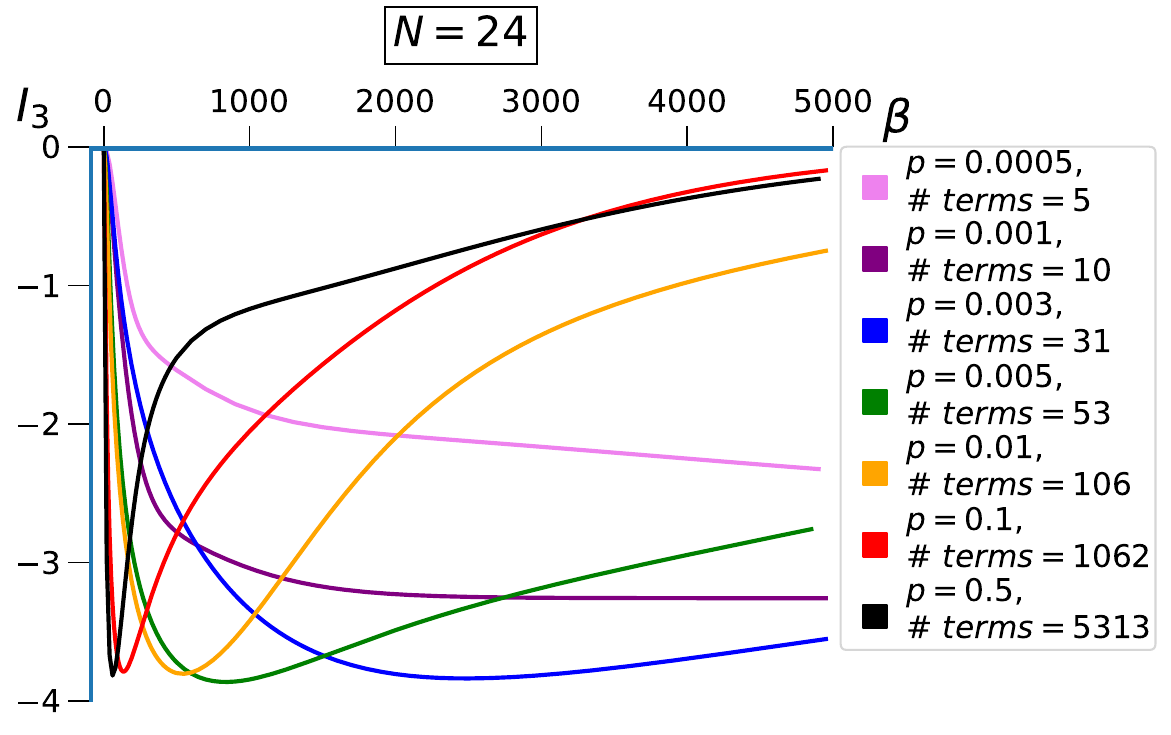} 
\caption{Tripartite information for the thermal state $\rho_{\beta}$ in the sparse SYK model with $N=12,18,24$. 
The curves represent the ensemble average $\sum_{1 \leq i \leq N_{\rm ens}} I_3(\rho_{\beta,i})/N_{\rm ens}$. We have taken $N_{\rm ens} =50,10,1$ respectively.}
\label{fig:I3-syk-beta}
\end{figure}

{\bf Numerical result.}
By exact diagonalization, we can explicitly build the thermal state $\rho_\beta$ and evaluate $I_3(\rho_\beta)$. For higher system sizes (N=18,24) we have followed the proccedure mentioned in \cite{sumilan} closely.
In Fig.~\ref{fig:I3-syk-beta} we display the tripartite information as a function of $\beta$ for fixed sparseness, controlled by the parameter $p$. The qualitative nature of the plots agrees with the above discussion. 
As already pointed out, for $\beta\ll 1$ we have $I_3 = \mathcal{O}(\beta^2)$.
In fact, the tripartite information is only mildly affected by the sparseness. This fact can be observed more clearly by looking at the tripartite information as a function of the purity $\tr(\rho_{\beta}^2)$, whose value is related to $\beta$.
\begin{figure}[h]
\centering
\includegraphics[scale=0.53]{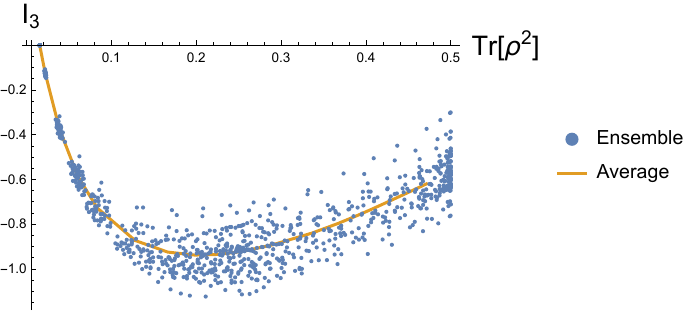} 
\qquad
\includegraphics[scale=0.6]{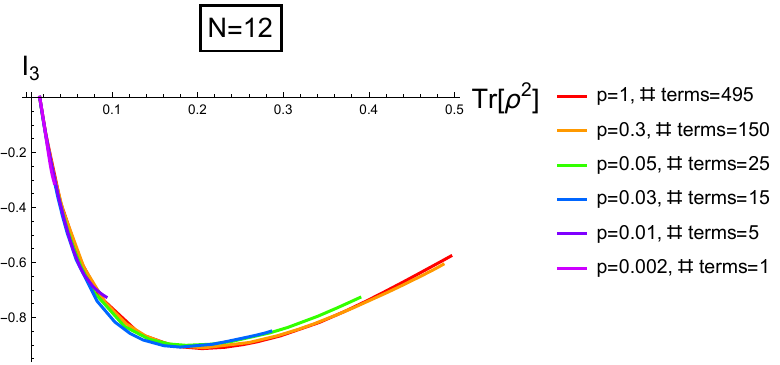} 
\qquad
\includegraphics[width=8cm,height=5cm]{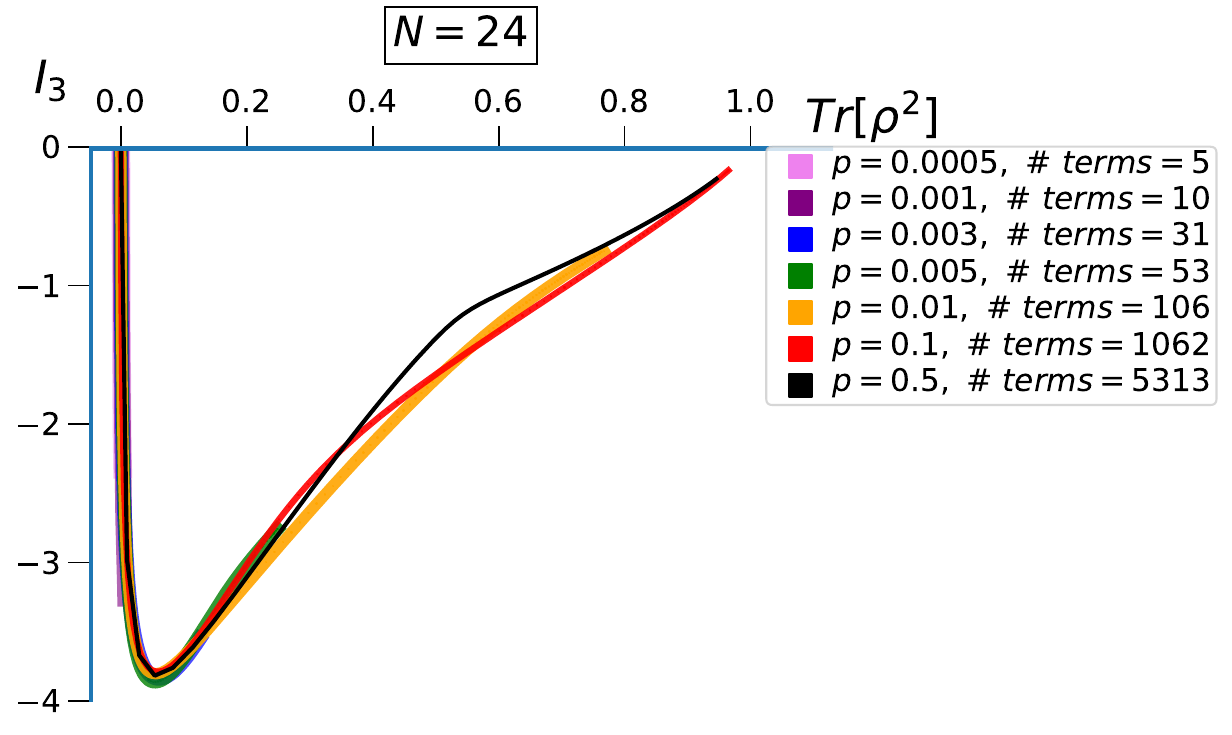}
\caption{Top: Tripartite information for the thermal state $\rho_{\beta}$ in the SYK model (left) and the sparse SYK model (right) with $N=12$. Bottom: Tripartite information for the thermal state $\rho_{\beta}$ in the sparse SYK model with $N=24$. Each point corresponds to an ensemble, while the curves represent the ensemble average $\sum_{1 \leq i \leq N_{\rm ens}} I_3(\rho_{\beta,i})/N_{\rm ens}$. We have taken $N_{\rm ens} =50,10,1$.}
\label{fig:I3-syk-purity}
\end{figure} 
At $\beta=0$, the state is maximally mixed, so the value of purity is minimum, $\tr(\rho_0^2) = d^{-1} = 2^{-n}$.
As $\beta$ increases, more and more energy eigenstates contributing to the thermal state in eq.~\eqref{eq:thermal-state} are suppressed. Thus, the purity monotonically increases with $\beta$. At $\beta = \infty$, the state is as close as possible to a pure state. Hence, the purity reaches its maximum value  $\tr(\rho_\infty^2) = g^{-1}$. 
Note that with the increasing of sparseness, the amount of symmetry in the Hamiltonian increases, and so does the degeneracy $g$ of the ground state. As a consequence, for increasing sparseness (decreasing $p$ in Fig.~\ref{fig:I3-syk-purity}) the value of purity at $\beta = \infty$ gets smaller. Therefore, the overall range of purity gets shortened.
Fig.~\ref{fig:I3-syk-purity} displays the tripartite information as a function of purity.
As it can be seen from the figure, by modifying the sparseness the curves almost overlap.
The only change in the tripartite information is the range of purity.
From Figs.~\ref{fig:I3-syk-beta} and \ref{fig:I3-syk-purity} it can be noted that the behavior of the tripartite information as a function of $\beta$ is qualitatively similar for $N=12,18,24$. Therefore, in the rest of the paper we mainly restrict our attention to $N=12$.

The most striking observation from the above plots is that the value of the tripartite information is always negative irrespective of the amount of sparseness, even when the effective number of terms is of $\order{(N)}$. In the literature, it has been argued that the sparse SYK model ceases to be holographic for sufficient sparseness when the effective number of terms $k N$ in the Hamiltonian becomes small enough, namely $k > 1/4$. The lower bound $N/4$ can be intuitively understood as follows. When the number of terms is less that ${N}/{4}$, some of the fermions are effectively decoupled from the rest (as the total number of flavors is $N$ and we are considering 4-fermion interactions). Thus, such a sparse model can no longer be holographic. However, in light of our result, in sparse SYK models the multipartite information is not enough to distinguish between the existence of a gravity dual or not. The negativity of the tripartite information can rather be seen as a necessary condition for the existence of a holographic dual.

In order to better illustrate how the nature of the ground states determines the sign of the tripartite information at large $\beta$, it will be useful to consider simpler models of fermi interactions without randomness. To this end, we now present two variations of fermi vector models with 4-fermi interactions without random couplings. 
\\

\subsection{Tripartite information for the vector model}
\label{subsec:tripartite-info-vector}

We will now show how the interaction pattern in the (sparse) SYK model is crucial for the mutual information of the thermal state to be monogamous. To this end, we consider vector models with the same degrees of freedom as the sparse SYK, but a simpler interacting pattern of the $N$-flavored Majorana fermions. As we will see, in this model the sign of the tripartite information for the thermal state depends on the choice of the subregions $A, B, C$.
\\

\begin{figure}[h]
\centering
\includegraphics[scale=0.45]{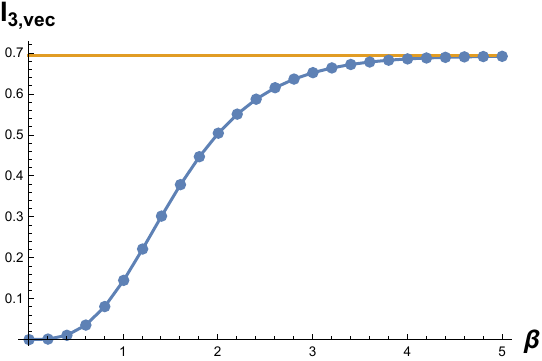} 
\qquad
\includegraphics[scale=0.45]{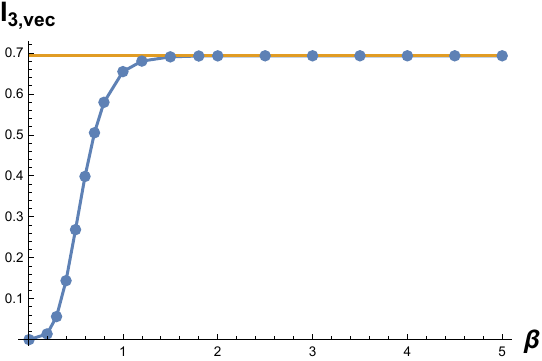} 
\qquad
\includegraphics[scale=0.45]{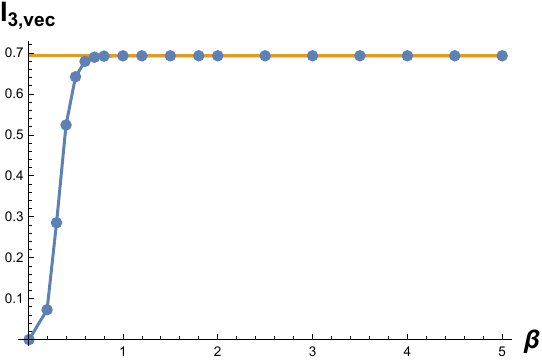} 
\caption{Tripartite information for the thermal state $\rho_\beta$ in the vector model \eqref{eq:H-vec1} with $N=6,12,18$.}
\label{fig:I3-vec1}
\end{figure}

{\bf Interaction of consecutive fermions.}
The vector model described by the Hamiltonian
\begin{align}
\label{eq:H-vec1}
H_{V1} = \left( \sum_{a=1}^{N/2} \psi_{2a-1} \psi_{2a} \right)^2
\end{align}
can be solved exactly.
\begin{center}
\begin{tikzpicture}
 \draw[fill=blue!40!white] (0,0) circle [radius=0.5];
 \node at (0,0) {$\psi_1$}; 
 \draw (0.5,0)--(1,0);
 \draw[fill=blue!40!white] (1.5,0) circle [radius=0.5];
 \node at (1.5,0) {$\psi_2$};
 \draw (-0.5,-0.6)--(-0.5,-0.7)--(2,-0.7)--(2,-0.6);
 \node[below] at (0.75,-0.7) {$A$}; 
 \draw[fill=blue!40!white] (3,0) circle [radius=0.5];
 \node at (3,0) {$\psi_3$};
 \draw (3.5,0)--(4,0);
 \draw[fill=blue!40!white] (4.5,0) circle [radius=0.5];
 \node at (4.5,0) {$\psi_4$};
 \draw (2.5,-0.6)--(2.5,-0.7)--(5,-0.7)--(5,-0.6);
 \node[below] at (3.75,-0.7) {$B$}; 
 \draw[fill=blue!40!white] (6,0) circle [radius=0.5];
 \node at (6,0) {$\psi_5$};
 \draw (6.5,0)--(7,0);
 \draw[fill=blue!40!white] (7.5,0) circle [radius=0.5];
 \node at (7.5,0) {$\psi_6$};
 \draw (5.5,-0.6)--(5.5,-0.7)--(8,-0.7)--(8,-0.6);
 \node[below] at (6.75,-0.7) {$C$}; 
 \end{tikzpicture}
\end{center}
In terms of Dirac fermions, the Hamiltonian \eqref{eq:H-vec1} can be expressed as
\begin{align}
H_{V1} = - \left( \frac{N}{4} - \sum_{a=1}^{N/2}  N_a  \right)^2 = - \left( \frac{N}{4} - Q \right)^2 \, ,
\end{align}
where $Q$ is the fermion charge operator defined in eq.~\eqref{eq:def-Q}.
The eigenstates of $H$ are simply the spin states $\ket{j}$ with eigenvalues
\begin{align}
H_{V1} \ket{j} = - \left( \frac{N}{4} - \# {\rm spin} \uparrow \right)^2 \ket{j} \, .
\end{align}
Clearly, there are two degenerate GSs $\ket{\uparrow \dots \uparrow}$ and $\ket{\downarrow \dots \downarrow}$, implying that
\begin{align}
\rho_\infty = \frac{1}{2} \left( \ketbra{\uparrow \dots \uparrow} + \ketbra{\downarrow \dots \downarrow} \right) \, .
\label{rinfv1}
\end{align}
Defining $A,B,C$ as in eq.~\eqref{eq:ABC-tripartite}, the correlations between $A,B$ and $A,C$, as well as the correlations between $A,BC$, are the same: $I_2(A,B) = I_2(A,C) = I_2(A,BC) = \log2$.
More explicitly, we have $S_A = S_B = S_C = \log 2$ and $S_{AB} = S_{AC} = S_{BC} = \log 2$.
Therefore $I_3(\rho_\infty) = \log2 >0$, the mutual information of the thermal state is not monogamous, see also the footnote \ref{footnote:I3}. 
In Fig. \ref{fig:I3-vec1} we display the tripartite information as a function of the inverse temperature $\beta$.
\\

\begin{figure}[h]
\centering
\includegraphics[scale=0.45]{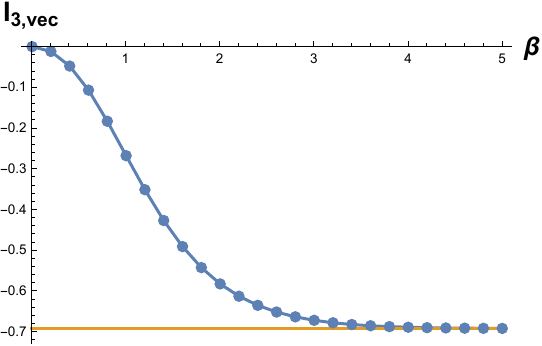} 
\qquad
\includegraphics[scale=0.45]{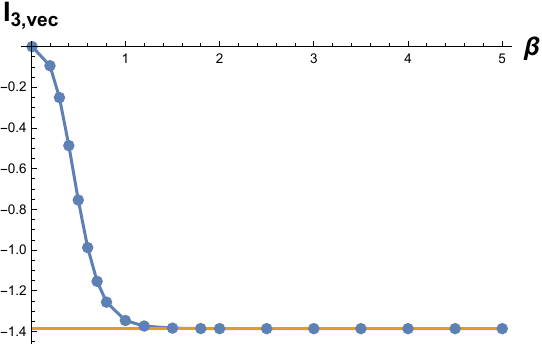} 
\qquad
\includegraphics[scale=0.45]{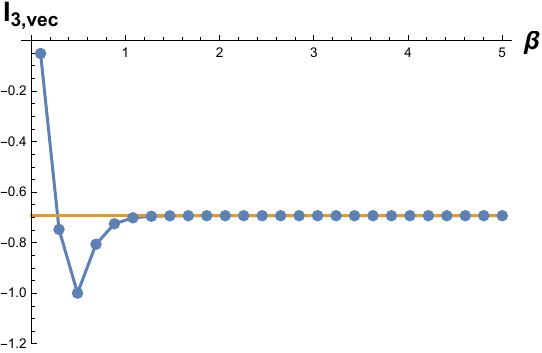} 
\caption{Tripartite information for the thermal state $\rho_\beta$ in the vector model \eqref{eq:H-vec2} with $N=6,12,18$.}
\label{fig:I3-vec2}
\end{figure}

{\bf Interaction of non-consecutive fermions.}
Let us now consider the following variation of the previous model
\begin{align}
\label{eq:H-vec2}
H_{V2} = \left( \sum_{a=1}^{N/2} \psi_{a} \psi_{N/2+a} \right)^2 \, .
\end{align}
\begin{center}
\begin{tikzpicture}
 \draw[fill=blue!40!white] (0,0) circle [radius=0.5];
 \node at (0,0) {$\psi_1$}; 
 \draw[fill=blue!40!white] (1.5,0) circle [radius=0.5];
 \node at (1.5,0) {$\psi_2$};
 \draw (-0.5,-0.6)--(-0.5,-0.7)--(2,-0.7)--(2,-0.6);
 \node[below] at (0.75,-0.7) {$A$}; 
 \draw[fill=blue!40!white] (3,0) circle [radius=0.5];
 \node at (3,0) {$\psi_3$};
 \draw[fill=blue!40!white] (4.5,0) circle [radius=0.5];
 \node at (4.5,0) {$\psi_4$};
 \draw (2.5,-0.6)--(2.5,-0.7)--(5,-0.7)--(5,-0.6);
 \node[below] at (3.75,-0.7) {$B$}; 
 \draw[fill=blue!40!white] (6,0) circle [radius=0.5];
 \node at (6,0) {$\psi_5$};
 \draw[fill=blue!40!white] (7.5,0) circle [radius=0.5];
 \node at (7.5,0) {$\psi_6$};
 \draw (5.5,-0.6)--(5.5,-0.7)--(8,-0.7)--(8,-0.6);
 \node[below] at (6.75,-0.7) {$C$};
 \draw (0,0.5)--(0,0.6)--(4.5,0.6)--(4.5,0.5);
 \draw (1.5,0.5)--(1.5,0.8)--(6,0.8)--(6,0.5);
 \draw (3,0.5)--(3,1)--(7.5,1)--(7.5,0.5);  
 \end{tikzpicture}
\end{center}
By relabeling the Majorana fermions, we can directly see that this model is equivalent to the previous one, so the spectrum is the same. However, the choice of subregions $A,B,C$ differs from the previous case.
Indeed, in the case at hand, Dirac fermions (or spins) belonging to different subregions interact non trivially.
As a result, the tripartite information turns out to be negative, see Fig.~\ref{fig:I3-vec2}.
To gain a better grasp on this result, let us consider the zero temperature state
\begin{align}
\rho_\infty = \frac{1}{2} \left( \ketbra{GS_1} + \ketbra{GS_2} \right) \, ,\label{rinfv2}
\end{align}
where the expressions of the ground states in the spin basis change with the number of Majorana fermions $N$.
Numerically, we have checked that in all the three cases shown in Fig.~\ref{fig:I3-vec2} we have
$I_2(A,B) = I_2(B,C) < 2 I_2(B,AC)$. In other words, the mutual information is here monogamous, see footnote~\ref{footnote:I3}.

	Comparing the two models above in the large $\beta$ regime, we see that the ground state density matrix in eq.~\eqref{rinfv1} and eq.~\eqref{rinfv2} are very different in terms of the ``partition" basis, even though there are only two states involved. We denote as ``partition" basis the spin basis given in eq.~\eqref{eq:spin-base} along with the partition defined in eq.~\eqref{eq:def-Aj}. In eq.~\eqref{rinfv1}, the density matrix has a contribution from only two particular orthonormal spin states, namely $ \ket{\uparrow \dots \uparrow}$ and $\ket{\downarrow \dots \downarrow}$. On the other hand, in eq.~\eqref{rinfv2}, the states $\ket{GS_1}$ and $\ket{GS_2}$ are more complicated linear combinations in terms of the ``partition" basis.\footnote{Of course, whether such linear combinations look ``complicated'' or not is a basis-dependent issue. Here we measure the  degree of complication in terms of the ``partition" basis.}
	An empirical observation that can be drawn from the above examples is that when more complicated states are involved for the density matrix in terms of the ``partition" basis, then the corresponding tripartite information will be negative.  Whether the density matrix is written in a complicated way or not in the natural basis for subsystem division determines the sign of $I_3$. 	We confirm this expectation numerically in Appendix \ref{negtrp}. However, we do not have a clear understanding of how much complication is necessary for $I_3$ to be negative or positive.

	Going back to the case of sparse SYK, which involves random couplings, it is easy to understand that the typical ground states in this model will be a linear combination of many orthonormal states in terms of the spin basis leading to a negative tripartite information at large $\beta$. On the other hand, if we choose a particular atypical ground state of sparse SYK models that involve only few orthonormal states in terms of the ``partition" basis, then we expect that the tripartite information will be positive. 



\section{Multipartite information in the sparse SYK}
\label{sec:multipartite-info}

In the previous section we have shown that in the sparse SYK model the tripartite information $I_3(A,B,C)$ of the thermal state $\rho_\beta$, with $A,B,C$ containing the same number of Majorana fermions, is always negative. 

We now generalize this quantity to an arbitrary number of partitions of the total system. 
To this purpose, let us orderly group the $n = N/2$ Dirac fermions into $M$ subregions $A_1, \dots, A_M$. Let us suppose that the subregions contain different numbers of fermions $n_j$, but each of the $n$ Dirac fermions is included in one and only one subregion.
Namely, $\cup_{i=1}^M A_i$ is the total system and $A_i \cap A_j = \emptyset$ for any $i \neq j$:
\begin{align}
A_j = \{ n_1 + \dots + n_{j-1} +1, \dots, n_1 + \dots + n_{j-1} + n_{j} \} \, ,
\qquad
\sum_{j=1}^{M} n_j = n \, .
\end{align}

The multipartite information $I_{\cal{N}}$, for ${\cal{N}} ( \le M)$ subregions out of $M$ total subregions, is defined as
\begin{align}
\label{eq:multi-def}
I_{\cal{N}} = \sum_{s} (-1)^{\left| s \right| +1} S(s) \, ,
\end{align}
where the sum runs over all possible subsets $s$ of 
$\{ A_{i_1}, \dots, A_{i_{\cal{N}}} \}$ and $\left| s \right|$ denotes the cardinality of the subset $s$. 

\subsection{Multipartite information for equi-partite system}
\label{subsec:equi-parties}

We here suppose that each subregion contains the same number of Dirac fermions. 
Then, in the large $n$ limit, due to the `democratic' interactions in the Hamiltonian in eq.~\eqref{eq:H-syk}, it is reasonable to assume that the entanglement entropy for the subset $s$ just depends on the cardinality $|s|$: $S(s) = S_{|s|}$. 
Consequently, the multipartite information \eqref{eq:multi-def} reads
\begin{align}
\label{eq:multi-democratic}
I_\mathcal{N} = \sum_{|s|=1}^{\mathcal{N}} (-1)^{|s|+1} \binom{\mathcal{N}}{|s|} S_{|s|} \, .
\end{align}
In this section, we focus on $\mathcal{N} = M$.
Based on this assumption and on the equi-partitioning of the system, we will be able to guess the behavior of the multipartite information $I_{\mathcal{N} =M}$.
However, this result does not generalize to the case $\mathcal{N} < M$ and to the case in which the subregions contain a different number of Dirac fermions. 
\\

{\bf Infinite temperature limit.}
The full system $A_1 \dots A_{\mathcal{N}}$ is in the maximally mixed state $\rho_0$.
Note that any partial trace of $\rho_0$ still results in a maximally mixed state of the given subsystem.
Therefore, we have $S_{|s|} = |s| n/\mathcal{N} \log2$, being $|s| n/\mathcal{N}$ the number of Dirac fermions in the subset with cardinality $|s|$. Then,
\begin{align}
I_\mathcal{N} (\rho_0) &= 
\frac{n \log2}{\mathcal{N}} \sum_{|s|=1}^{\mathcal{N}} (-1)^{|s|+1} \binom{\mathcal{N}}{|s|} |s| \\ \nonumber
&= \frac{n \log2}{\mathcal{N}} \left[ \frac{d}{dx} \sum_{|s|=0}^{\mathcal{N}} (-1)^{|s|+1} \binom{\mathcal{N}}{|s|} x^{|s|} \right]_{x=1}
= - \frac{n \log2}{\mathcal{N}} \left[ \frac{d}{dx} \left( 1-x \right)^{\mathcal{N}} \right]_{x=1} =0 \, .
\end{align}
\\

\begin{figure}[h]
\centering
\includegraphics[scale=0.7]{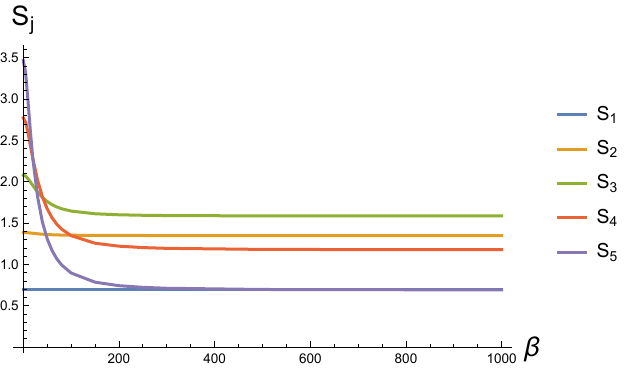} 
\caption{Entanglement entropy $S_j = S(A_1 \dots A_j)$ as a function of $\beta$ for subsets with different cardinality $j$. We have split the total system of $n=5$ Dirac fermions into $\mathcal{N}=5$ subregions, each of which containing one fermion. The figure shows the result averaged over $N_{\rm ens}=100$ ensembles for the SYK model ($p=1$).}
\label{fig:Sj-beta}
\end{figure}

{\bf Large but finite temperature limit.}
When $\beta \to 0^+$, the state $\rho_{0^+}$ of the full system $A_1 \dots A_{\mathcal{N}}$ slightly deviates from the maximally mixed state $\rho_0$. 
Consequently, the values of all the entropies $S_{|s|}$ in eq.~\eqref{eq:multi-democratic} change from their maximal values. This leads to $I_{\mathcal{N}}(\rho_{0^+})$ being non-zero.
As exemplified in Fig. \ref{fig:Sj-beta}, the higher the subset cardinality $|s|$, the faster the decreasing of $S_{|s|}$ as a function of $\beta$. 
Intuitively, this is due to the fact that the perturbation of the maximally mixed state due to the decreasing temperature is washed out more and more as we keep tracing out more degrees of freedom.
Therefore, the leading contribution to $I_{\mathcal{N}}(\rho_{0^+})$ is given by the last term in eq.~\eqref{eq:multi-democratic}:
\begin{align}
{\rm sign} \left( I_{\mathcal{N}}(\rho_{0^+}) \right) = -(-1)^{\mathcal{N}+1} = (-1)^{\mathcal{N}} \, .
\label{eq:IN-zero-plus}
\end{align}
\\

{\bf Zero temperature limit.}
The full system $A_1 \dots A_{\mathcal{N}}$ is in the state $\rho_\infty$, to which just the GSs contribute.
Let us suppose that the GS is non-degenerate, a situation which can be achieved for $N \, {\rm mod} \, 8 \neq 4$ by selecting eigenstates with fixed chirality. Then, $S_{\mathcal{N}} =0$ and $S_{\mathcal{N}-j} = S_j$ for any $j=1, \dots \mathcal{N}-1$ (see Fig. \ref{fig:entropy-GS}). So, the binomial factor with $\alpha = \mathcal{N}$ in eq.~\eqref{eq:multi-democratic} and the entanglement entropy $S_{|s|}$ are both invariant under $|s| \to \mathcal{N}-|s|$. 
Let us distinguish between two cases:
\begin{itemize}
 \item {\bf $\mathcal{N}$ odd.} 
 The $(\mathcal{N}-j)$-th term in eq.~\eqref{eq:multi-democratic} cancels the $j$-th term, leading to $I_{\mathcal{N}} (\rho_{\infty, g=1}) = 0$. This agrees with the observation that $I_3 =0$ for a pure state.
 \item {\bf $\mathcal{N}$ even.}
 The multipartite information reads
  \begin{align}
 I_{\mathcal{N}} (\rho_{\infty, g=1}) = 2 \sum_{|s|=1}^{\left[ \frac{\mathcal{N}-1}{2} \right]} (-1)^{|s|+1} \binom{\mathcal{N}}{|s|} S_{|s|} + (-1)^{\frac{\mathcal{N}}{2}+1} \binom{\mathcal{N}}{\mathcal{N}/2} S_{\mathcal{N}/2}  \, ,
 \end{align}
 where $[.]$ denotes the integer part. Since the binomial factor grows polynomially in $|s|$ and the entropy grows at most linearly in $|s|$, the last term dominates, giving ${\rm sign} \left( I_{\mathcal{N}} (\rho_{\infty, g=1}) \right) = (-1)^{\mathcal{N}/2+1}$. 
\end{itemize}
In the presence of degeneracy, the $g$ GSs contribute to $\rho_\infty$. 
In this case, $S_{\mathcal{N}} = \log g$ and $S_{\mathcal{N}-j} \neq S_j$,
so the multipartite information is non-vanishing for odd $\mathcal{N}$ too.
However, due to the symmetry of the binomial factor under $|s| \to \mathcal{N}-|s|$,
by an argument analogous to the previous one, we expect\footnote{Note that when $\mathcal{N}$ is odd, the binomial factor acquires its maximum value at both $|s_1| = (\mathcal{N} -1)/2$ and $|s_2| = (\mathcal{N} +1)/2$. If the state is not pure $S_{|s_1|} \neq S_{|s_2|}$, but since $|s_1| < |s_2|$ we expect $S_{|s_1|} \lesssim S_{|s_2|}$. Thus, the term $s=|s_2|$ dominates.}
\begin{align}
\label{eq:IN-infinity}
{\rm sign} \left( I_{\mathcal{N}} (\rho_{\infty, g}) \right) = (-1)^{\left[ \frac{\mathcal{N}+1}{2} \right] +1} \, .
\end{align}
\\

{\bf Finite temperature.}
From eqs.~\eqref{eq:IN-zero-plus} and \eqref{eq:IN-infinity},
the mutual information $I_2$ is positive for both $\beta = 0^+$ and $\beta \to +\infty$.
This agrees with the known fact that $I_2 \geq 0$ for any $\beta$.
Similarly, the tripartite information $I_3$ is negative for both $\beta = 0^+$ and $\beta \to +\infty$.
In Sec. \ref{subsec:tripartite-info-syk} we have discussed that numerical results suggest indeed $I_3 \leq 0$ for any $\beta$.

Let us discuss the result for higher number of parties $\mathcal{N} \geq 4$.
For a four-partite system, $I_4(\rho_{0^+}) \geq 0$ and $I_4(\rho_{\infty,g}) \leq 0$, so $I_4(\rho_\beta)$ must vanish at least for one finite value of $\beta$. Even though we are not able to get this exact value, we can expect the vanishing of $I_4(\rho_\beta)$ to happen at $\beta\ll1$. 
As we have argued before, the term $|s| = \mathcal{N}$ in eq.~\eqref{eq:multi-democratic} is responsible for the sign of $I_4(\rho_{0^+})$. As it is clear from Fig. \ref{fig:Sj-beta}, the entanglement entropy $S_\mathcal{N}$ decreases fast with $\beta$, so that at small $\beta$ the term $|s| = \mathcal{N}-1$, with opposite sign, competes. This effect causes the vanishing of $I_4(\rho_\beta)$ at $\beta \ll1$.
\\
The same argument, modulo a flip of signs, applies to the five-partite case $I_5(\rho_\beta)$.
\\
Instead, for a six-partite system, $I_6(\rho_{0^+}) \geq 0$ and $I_6(\rho_{\infty,g}) \geq 0$. 
Since the above mechanism still takes place, $I_6(\rho_\beta)$ must have at least two zeroes.
\\
Generalizing this reasoning, we conclude that
\begin{align}
\# {\rm zeroes \, of} \, I_\mathcal{N}(\rho_\beta) \geq \left[ \frac{\mathcal{N}-2}{2} \right] \, .
\end{align}
As we will see from the numerical results, the bound on the number of zeroes is saturated by the full SYK model ($p=1$). 
\\

\begin{figure}[h]
\centering
\includegraphics[scale=0.6]{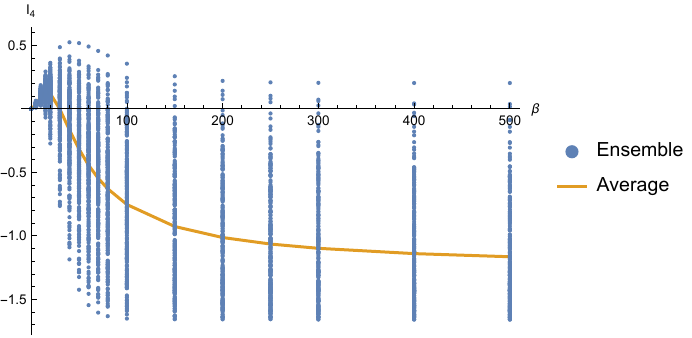} 
\qquad
\includegraphics[scale=0.6]{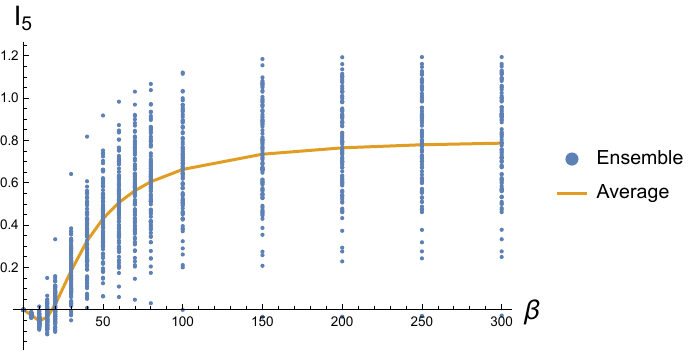} 
\qquad
\includegraphics[scale=0.6]{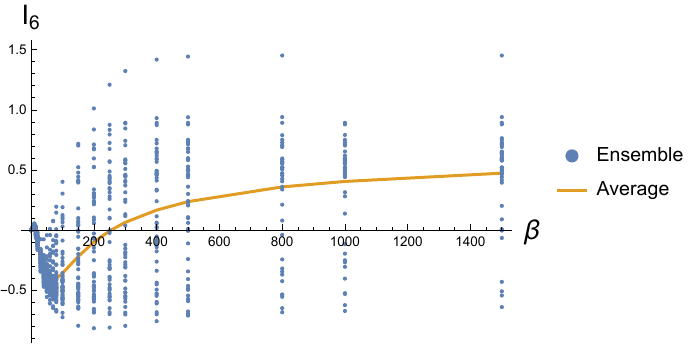} 
\qquad
\includegraphics[scale=0.6]{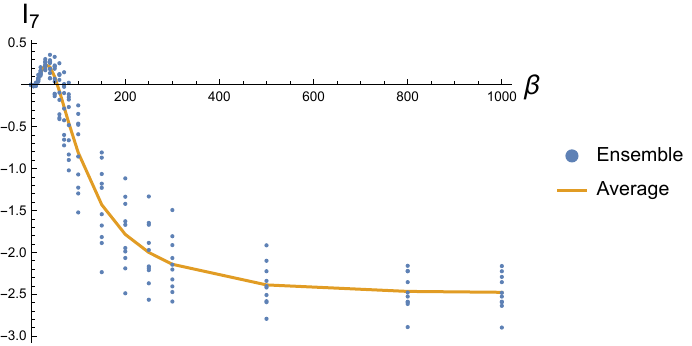} 
\caption{Multipartite information for the thermal state $\rho_{\beta}$ in the SYK model with $N=8,10,12,14$. Each point corresponds to an ensemble, while the curves represent the ensemble average $\sum_{1 \leq i \leq N_{\rm ens}} I_\mathcal{N}(\rho_{\beta,i})/N_{\rm ens}$. We have taken $N_{\rm ens} =300,100,50,10$ respectively.}
\label{fig:multi-syk-beta}
\end{figure}

\begin{figure}[h]
\centering
\includegraphics[scale=0.6]{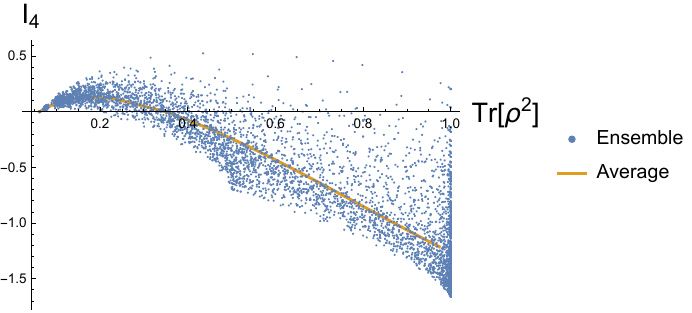} 
\qquad
\includegraphics[scale=0.6]{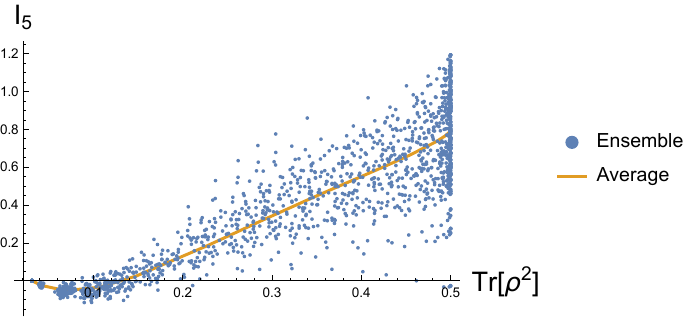} 
\qquad
\includegraphics[scale=0.6]{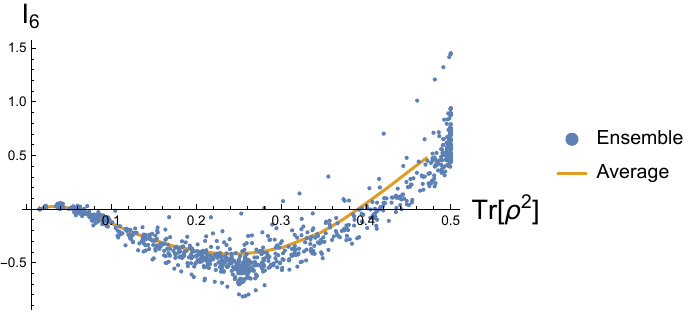} 
\qquad
\includegraphics[scale=0.6]{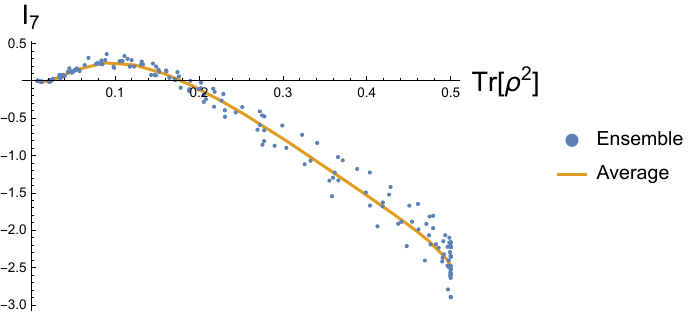} 
\caption{Multipartite information for the thermal state $\rho_{\beta}$ in the SYK model with $N=8,10,12,14$. Each point corresponds to an ensemble, while the curves represent the ensemble average $\sum_{1 \leq i \leq N_{\rm ens}} I_\mathcal{N}(\rho_{\beta,i})/N_{\rm ens}$. We have taken $N_{\rm ens} =300,100,50,10$ respectively.}
\label{fig:multi-syk-pur}
\end{figure}

\begin{figure}[h]
\centering
\includegraphics[scale=0.5]{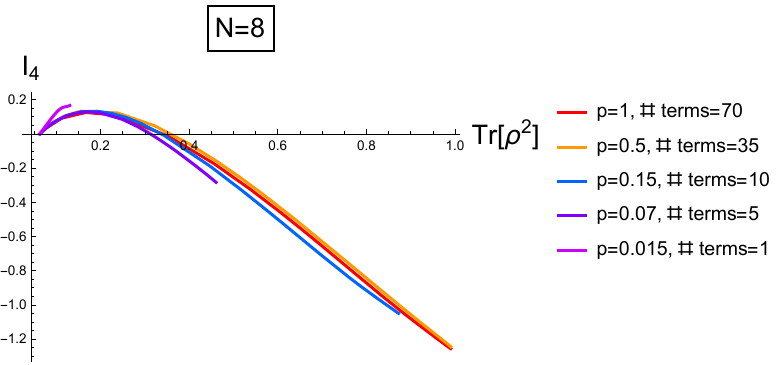} 
\qquad
\includegraphics[scale=0.5]{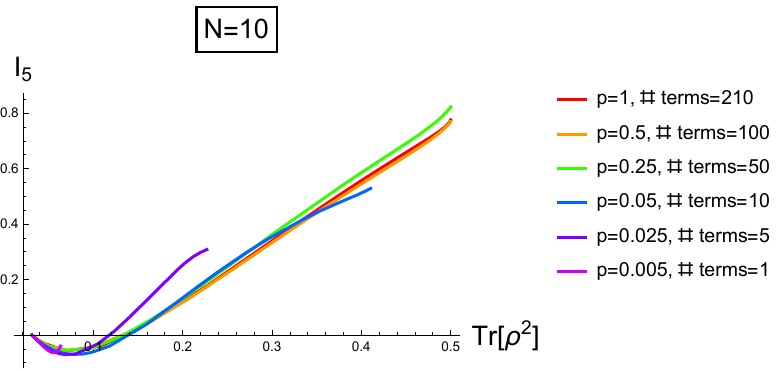} 
\qquad
\includegraphics[scale=0.5]{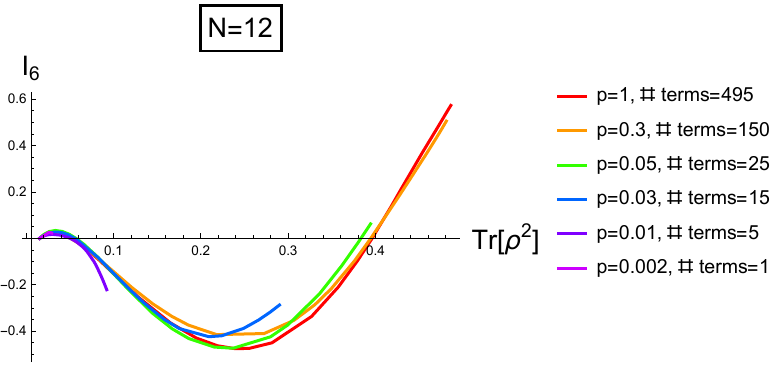} 
\caption{Multipartite information for the thermal state $\rho_{\beta}$ in the sparse SYK model with $N=8,10,12$. Each curve represent the ensemble average $\sum_{1 \leq i \leq N_{\rm ens}} I_\mathcal{N}(\rho_{\beta,i})/N_{\rm ens}$ for a fixed value of the sparseness $p$. For a fixed number of Majorana fermions $N$, we have taken $N_{\rm ens} =300,100,50$ respectively.}
\label{fig:multi-sparse-pur}
\end{figure}

{\bf Numerical results.}
We display numerical results of $I_{\mathcal{N}}(\rho_\beta)$ for full SYK ($p=1$) in Fig. \ref{fig:multi-syk-beta}.
\\
Based on the numerical analysis and in line with the previous discussion, the following properties of $I_{\mathcal{N}}(\rho_\beta)$ for equi-partitioning of the full system can be deduced:
\begin{enumerate}
 \item $I_{\mathcal{N}}(\rho_0) =0$.
 \item Conjecture: ${\rm sign} \left( I_{\mathcal{N}}(\rho_{0^+}) \right) = (-1)^{\mathcal{N}}$. 
 \item Conjecture: ${\rm sign} \left( I_{\mathcal{N}} (\rho_{\infty, g}) \right) = (-1)^{\left[ \frac{\mathcal{N}+1}{2} \right] +1}$.
 \item Conjecture: $\# {\rm zeroes \, of} \, I_\mathcal{N}(\rho_\beta) \geq \left[ \frac{\mathcal{N}-2}{2} \right]$.
\end{enumerate}
The numerical results suggest that the bound $4$ is saturated.
The corresponding plots in terms of the purity $\tr ( \rho_\beta^2 )$ are also shown, see Fig. \ref{fig:multi-syk-pur}.
As can be noticed from the figure, the same properties hold true for the multipartite information of full SYK as a function of purity.
As mentioned in section \ref{sec:tripartite-info} in the discussion leading to Fig.~\ref{fig:I3-syk-purity}, with the increasing of sparseness the range of purity decreases. 
Consequently, properties 3 and 4 are in general not valid in the sparse SYK model for small enough $p$. 
As can be noticed from Fig. \ref{fig:multi-sparse-pur}, the narrowing of the range of purity is the main effect of sparseness.


\section{Multipartite information inequalities}
\label{sec:EE-inequalities}

From now on, we relax the condition of equi-partitioning of the system and we take $\mathcal{N} \leq M$.
In QFTs, the quantities $I_{\mathcal{N} \leq 4}$ do not provide any information about holography, having no definite sign in holographic field theories.
As we have seen in Sec. \ref{sec:multipartite-info}, the same property is shared by the multipartite information $I_{\mathcal{N} \leq 4}$ of the thermal state in the (sparse) SYK model as a function of temperature (or, equivalently, purity).
In \cite{Bao:2015bfa}, it has been shown that in holographic QFTs specific combinations of entanglement entropies, which can be rearranged in specific combinations of the multipartite information quantities in eq.~\eqref{eq:multi-def}, have instead a definite sign.
Whereas for $\mathcal{N}=4$ partitions (the uplift of) $I_2 \geq 0$ and $I_3 \leq 0$ are the only independent inequalities, for $\mathcal{N} \geq 5$ partitions new holographic inequalities have been unraveled. 
More specifically, for $\mathcal{N} = 5$ there are five new classes of inequalities. One representative of each class is written in Table 1 of \cite{Hernandez-Cuenca:2023iqh}. For convenience, we report here such representatives:
\begin{align}
\label{eq:W1}
W_1 &= I_{ABCD} + I_{ACDE} - I_{ACD} - I_{ACE} - I_{BCD} \geq 0 \, ,\\
\label{eq:W2}
W_2 &= I_{ABCD} + I_{BCDE} - I_{ABE} - I_{ACD} - I_{BCD} - I_{CDE} \geq 0 \, ,\\
\label{eq:W3}
W_3 &= I_{ABCD} + I_{ACDE} + I_{BCDE} - I_{ACD} - I_{ACE} - I_{BCD} - I_{BDE} \geq 0 \, ,\\
\label{eq:W4}
W_4 &= I_{ABCE} + I_{ACDE} + I_{BCDE} - I_{ABD} - I_{ACE} - I_{BCE} - I_{CDE} \geq 0 \, ,\\
\label{eq:W5}
W_5 &= 2 I_{ABCD} + I_{ABCE} + 2 I_{ABDE} + I_{ACDE}  \nonumber \\
 &  \qquad -I_{ABD} - 2 I_{ABE} - 2 I_{ACD} - I_{ACE} - I_{BCD} - I_{BDE} \geq 0 \, ,
\end{align} 
where $I_{ABCD} \equiv I_4 (A,B,C,D)$ and $I_{ABC} \equiv I_3 (A,B,C)$.
In the first class, the independent inequalities are obtained by proper permutations of the five subregions $A,B,C,D,E$ (avoiding the permutations which leave $W_1$ invariant), together with the substitution of $C$ with the purifier, followed by all the possible permutations of $A,B,C,D,E$.
Similarly, in the second class the independent inequalities are given by the permutations of the five subregions (except those which leave $W_2$ invariant), in addition to the substitution of $A,C,D,$ or $E$ with the purifier, followed by all the permutations of $A,B,C,D,E$.
On the other hand, in the remaining three classes the substitution of one of the five subregions with the purifier is equivalent to a proper permutation of the five subregions. Therefore, in such three classes, all the independent inequalities are obtained by simply permuting the five subregions $A,B,C,D,E$ and ignoring the purifier.
See Appendix \ref{app:inequalities} for a more detailed analysis of the complete set of independent inequalities.

In this section, we check whether the inequalities in eqs.~\eqref{eq:W1}-\eqref{eq:W5} for the entanglement of flavors are satisfied in fermionic systems.
Note that the multipartite information quantities $I_{ABC(D)}$ in the inequalities do not involve the whole systems, i.e. $n_A + n_B + n_C ( + n_D) < n$ (in the terminology introduced in the previous section, $\mathcal{N} < M$). In general, we may choose the $\mathcal{N}=5$ partitions so that some of the spins are left out: 
\begin{center}
\begin{tikzpicture}
 \draw[fill=blue!40!white] (0,0) circle [radius=0.5];
 \draw[-{Latex[length=2mm,width=2mm]}] (0,-0.6) to  (0,0.8);
 \node at (1,0) {$\dots$};
 \draw[fill=blue!40!white] (2,0) circle [radius=0.5];
 \draw[-{Latex[length=2mm,width=2mm]}] (2,-0.6) to  (2,0.8);
 \draw (-0.5,-0.8)--(-0.5,-0.9)--(2.5,-0.9)--(2.5,-0.8);
 \node[below] at (1,-0.9) {$A=n_A$ spins}; 
 \node at (3,0) {$\dots$};
 \draw[fill=blue!40!white] (4,0) circle [radius=0.5];
 \draw[-{Latex[length=2mm,width=2mm]}] (4,0.6) to (4,-0.8);
 \node at (5,0) {$\dots$};
 \draw[fill=blue!40!white] (6,0) circle [radius=0.5];
 \draw[-{Latex[length=2mm,width=2mm]}] (6,-0.6) to  (6,0.8);
 \draw (3.5,-0.8)--(3.5,-0.9)--(6.5,-0.9)--(6.5,-0.8);
 \node[below] at (5,-0.9) {$E=n_E$ spins}; 
 \draw[fill=blue!40!white] (7.5,0) circle [radius=0.5];
 \draw[-{Latex[length=2mm,width=2mm]}] (7.5,-0.6) to  (7.5,0.8);
 \node at (8.5,0) {$\dots$};
 \draw[fill=blue!40!white] (9.5,0) circle [radius=0.5];
 \draw[-{Latex[length=2mm,width=2mm]}] (9.5,0.6) to (9.5,-0.8);
 \end{tikzpicture}
\end{center}
Despite this, we still take the full system of $n$ spins to be in the thermal state $\rho_\beta$.
We again stress that this configuration differs from the one we considered in the previous section.
In particular, the general form of $I_{\mathcal N =M}$ for equi-partite system shown in Fig.~\ref{fig:multi-sparse-pur} does not apply to the $I_{ABC(D)}$ considered here.
\\

\subsection{Multipartite information for the sparse SYK}
\label{subsec:multipartite-info-syk}

\begin{figure}[h]
\centering
\includegraphics[scale=0.6]{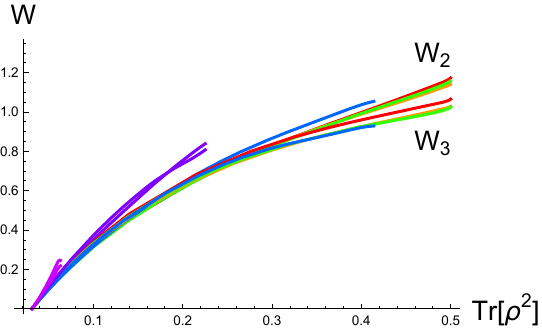} 
\qquad
\includegraphics[scale=0.6]{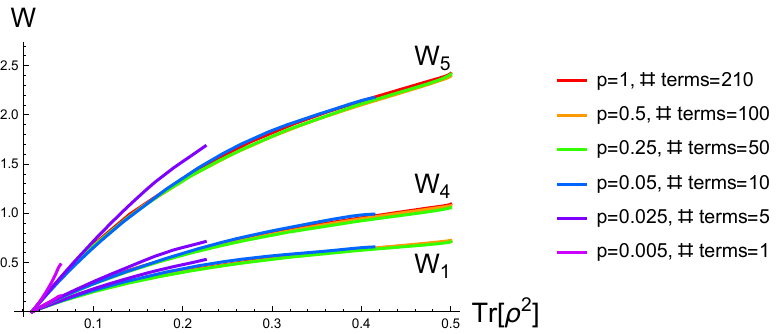}
\caption{Multipartite quantities in eqs.~\eqref{eq:W1}-\eqref{eq:W5} for the thermal state $\rho_{\beta}$ in the sparse SYK model with $N=10$. The curves represent the ensemble average $\sum_{1 \leq i \leq N_{\rm ens}} W_j(\rho_{\beta,i})/N_{\rm ens}$. We have taken $N_{\rm ens} =100$.}
\label{fig:HEI-syk1}
\end{figure}

\begin{figure}[h]
\centering
\includegraphics[scale=0.6]{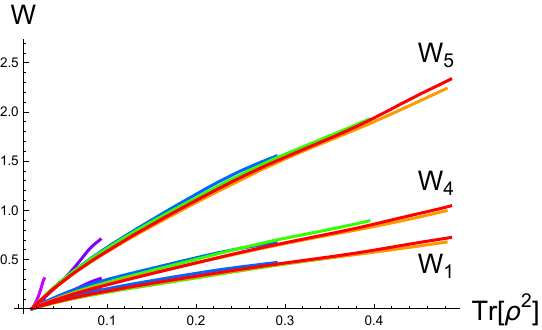}
\qquad
\includegraphics[scale=0.6]{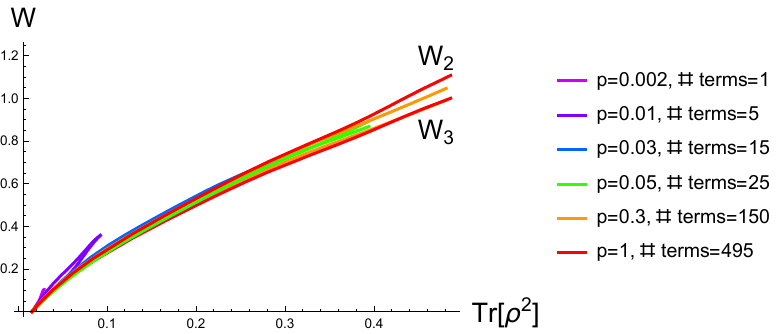} 
\qquad
\includegraphics[scale=0.6]{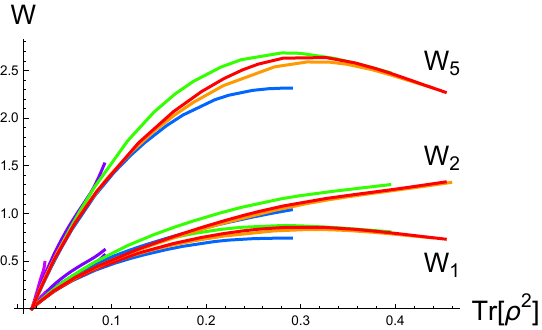}
\qquad
\includegraphics[scale=0.6]{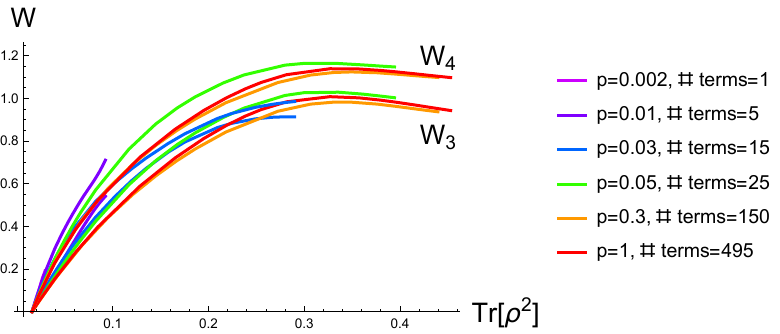}
\caption{Multipartite quantities in eqs.~\eqref{eq:W1}-\eqref{eq:W5} for the thermal state $\rho_{\beta}$ in the sparse SYK model with $N=12$. In the top plots we have taken $A,B,C,D,E$ to contain one spin each, while in the bottom plot $A$ contains two spins and $B,C,D,E$ one spin each. The curves represent the ensemble average $\sum_{1 \leq i \leq N_{\rm ens}} W_j(\rho_{\beta,i})/N_{\rm ens}$. We have taken $N_{\rm ens} =50$.}
\label{fig:HEI-syk2}
\end{figure}

We now show the quantities $W_i$ for the sparse SYK model as functions of the purity for different sparseness.
In this section, for illustrative purposes, we just pick one representative for each class of inequalities and display the effect of sparseness. The full set of inequalities for $\mathcal{N}=5$ partitions is shown in Appendix \ref{app:inequalities} for the full SYK. Qualitatively, the effect of sparseness on all such instances is the same as we will discuss in the present section.
In Fig.~\ref{fig:HEI-syk1}, we display the result for $N=10$ Majorana fermions, in which case we can just take each of the five partitions to contain a single spin.
In Fig.~\ref{fig:HEI-syk2}, we instead consider $N=12$. In the top of the figure we again take each of the five partitions to consist of a single spin, thus leaving one spin out of the partitions, while in the bottom of the figure we take one subregion to contain two spins. 
All the inequalities in eqs.~\eqref{eq:W1}-\eqref{eq:W5} are satisfied for every choice of the sparseness. 
Indeed, as for the multipartite information, the value of $W_i$ mostly depend on the purity of the thermal state $\rho_{\beta}$ rather than on the degree of sparseness of the model.
We have checked all the independent inequalities for $N=10,12$ and any possible choice of $3 \leq \mathcal{N} \leq 5$ partitions, see Appendix \ref{app:inequalities} for further details. All the inequalities turn out to hold for any $\beta$ and $p$.
\\

\subsection{Multipartite information for the vector model}
\label{subsec:multipartite-info-vector}

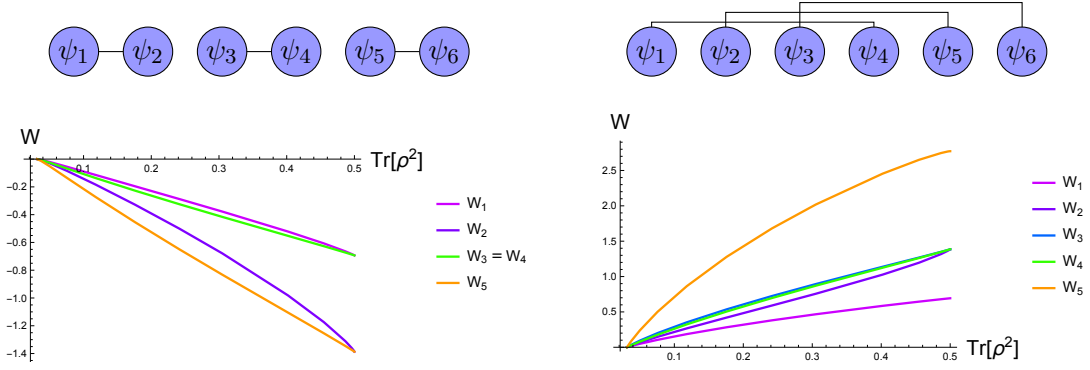
\begin{figure}[h]
\centering
\begin{tikzpicture}[scale=0.65]
 \draw[fill=blue!40!white] (0,0) circle [radius=0.5];
 \node at (0,0) {$\psi_1$}; 
 \draw (0.5,0)--(1,0);
 \draw[fill=blue!40!white] (1.5,0) circle [radius=0.5];
 \node at (1.5,0) {$\psi_2$};
 \node[below] at (0.75,-0.7) {}; 
 \draw[fill=blue!40!white] (3,0) circle [radius=0.5];
 \node at (3,0) {$\psi_3$};
 \draw (3.5,0)--(4,0);
 \draw[fill=blue!40!white] (4.5,0) circle [radius=0.5];
 \node at (4.5,0) {$\psi_4$};
 \draw[fill=blue!40!white] (6,0) circle [radius=0.5];
 \node at (6,0) {$\psi_5$};
 \draw (6.5,0)--(7,0);
 \draw[fill=blue!40!white] (7.5,0) circle [radius=0.5];
 \node at (7.5,0) {$\psi_6$};
 \end{tikzpicture}
 \qquad \qquad \quad
 \begin{tikzpicture}[scale=0.65]
 \draw[fill=blue!40!white] (0,0) circle [radius=0.5];
 \node at (0,0) {$\psi_1$}; 
 \draw[fill=blue!40!white] (1.5,0) circle [radius=0.5];
 \node at (1.5,0) {$\psi_2$};
 \node[below] at (0.75,-0.7) {}; 
 \draw[fill=blue!40!white] (3,0) circle [radius=0.5];
 \node at (3,0) {$\psi_3$};
 \draw[fill=blue!40!white] (4.5,0) circle [radius=0.5];
 \node at (4.5,0) {$\psi_4$};
 \draw[fill=blue!40!white] (6,0) circle [radius=0.5];
 \node at (6,0) {$\psi_5$};
 \draw[fill=blue!40!white] (7.5,0) circle [radius=0.5];
 \node at (7.5,0) {$\psi_6$};
 \draw (0,0.5)--(0,0.6)--(4.5,0.6)--(4.5,0.5);
 \draw (1.5,0.5)--(1.5,0.8)--(6,0.8)--(6,0.5);
 \draw (3,0.5)--(3,1)--(7.5,1)--(7.5,0.5);  
 \end{tikzpicture}
 \qquad
\includegraphics[scale=0.6]{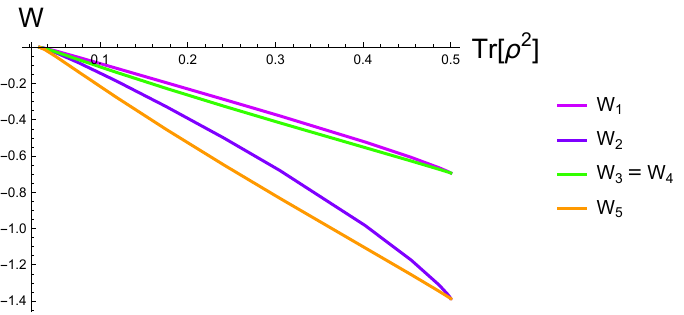} 
\qquad
\includegraphics[scale=0.6]{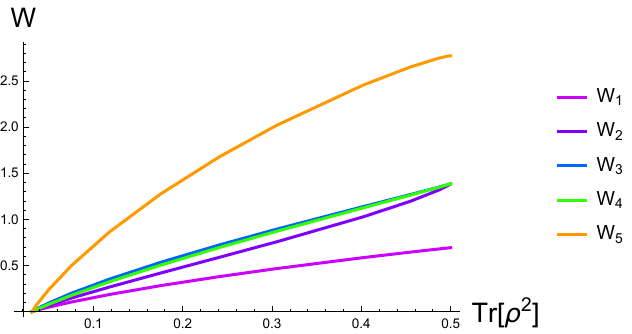} 
\caption{Multipartite quantities in eqs.~\eqref{eq:W1}-\eqref{eq:W5} for the thermal state $\rho_{\beta}$ in the vector model for consecutive (on the left) and non-consecutive (on the right) interacting fermions with $N=10$.}
\label{fig:HEI-vector1}
\end{figure}

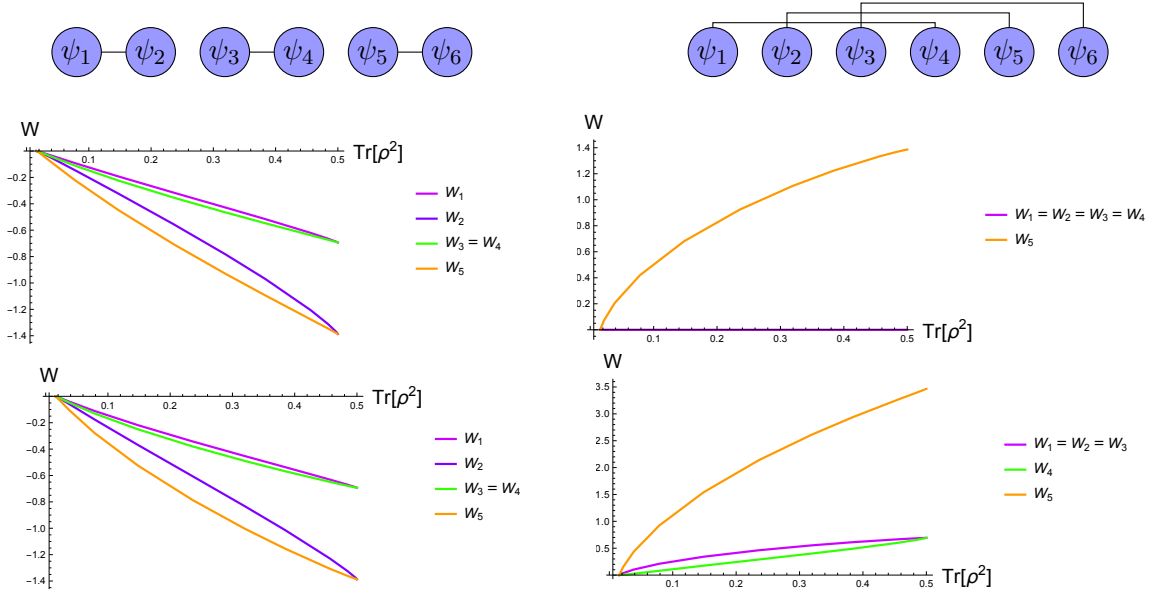
\begin{figure}[h]
\centering
\begin{tikzpicture}[scale=0.65]
 \draw[fill=blue!40!white] (0,0) circle [radius=0.5];
 \node at (0,0) {$\psi_1$}; 
 \draw (0.5,0)--(1,0);
 \draw[fill=blue!40!white] (1.5,0) circle [radius=0.5];
 \node at (1.5,0) {$\psi_2$};
 \node[below] at (0.75,-0.7) {}; 
 \draw[fill=blue!40!white] (3,0) circle [radius=0.5];
 \node at (3,0) {$\psi_3$};
 \draw (3.5,0)--(4,0);
 \draw[fill=blue!40!white] (4.5,0) circle [radius=0.5];
 \node at (4.5,0) {$\psi_4$};
 \draw[fill=blue!40!white] (6,0) circle [radius=0.5];
 \node at (6,0) {$\psi_5$};
 \draw (6.5,0)--(7,0);
 \draw[fill=blue!40!white] (7.5,0) circle [radius=0.5];
 \node at (7.5,0) {$\psi_6$};
 \end{tikzpicture}
 \qquad \qquad \qquad \quad
 \begin{tikzpicture}[scale=0.65]
 \draw[fill=blue!40!white] (0,0) circle [radius=0.5];
 \node at (0,0) {$\psi_1$}; 
 \draw[fill=blue!40!white] (1.5,0) circle [radius=0.5];
 \node at (1.5,0) {$\psi_2$};
 \node[below] at (0.75,-0.7) {}; 
 \draw[fill=blue!40!white] (3,0) circle [radius=0.5];
 \node at (3,0) {$\psi_3$};
 \draw[fill=blue!40!white] (4.5,0) circle [radius=0.5];
 \node at (4.5,0) {$\psi_4$};
 \draw[fill=blue!40!white] (6,0) circle [radius=0.5];
 \node at (6,0) {$\psi_5$};
 \draw[fill=blue!40!white] (7.5,0) circle [radius=0.5];
 \node at (7.5,0) {$\psi_6$};
 \draw (0,0.5)--(0,0.6)--(4.5,0.6)--(4.5,0.5);
 \draw (1.5,0.5)--(1.5,0.8)--(6,0.8)--(6,0.5);
 \draw (3,0.5)--(3,1)--(7.5,1)--(7.5,0.5);  
 \end{tikzpicture}
 \qquad
\includegraphics[scale=0.57]{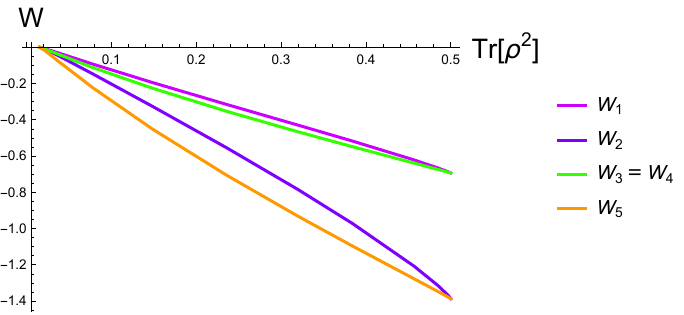} 
\qquad
\includegraphics[scale=0.57]{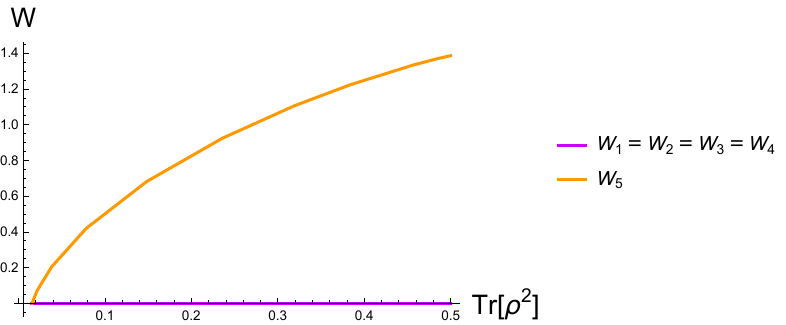} 
 \qquad
\includegraphics[scale=0.57]{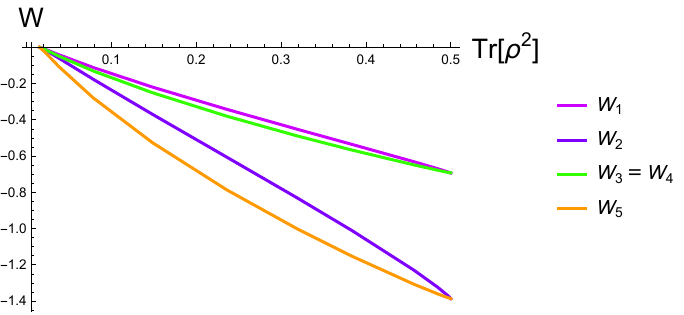} 
\qquad
\includegraphics[scale=0.57]{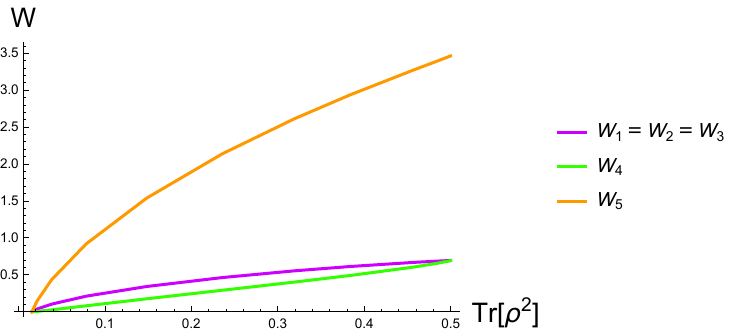} 
\caption{Multipartite quantities in eqs.~\eqref{eq:W1}-\eqref{eq:W5} for the thermal state $\rho_{\beta}$ in the vector model for consecutive (on the left) and non-consecutive (on the right) interacting fermions with $N=12$.
In the top plots we have taken $A,B,C,D,E$ to contain one spin each, while in the bottom plot $A$ contains two spins and $B,C,D,E$ one spin each.}
\label{fig:HEI-vector2}
\end{figure}

In Figs.~\ref{fig:HEI-vector1} and \ref{fig:HEI-vector2} we display the quantities $W_i$ for the vector model with consecutive fermions interacting (on the left) and non-consecutive fermions interacting (on the right).
Consistently with the result observed in Sec.~\ref{subsec:tripartite-info-vector},
all the inequalities in eqs.~\eqref{eq:W1}-\eqref{eq:W5} are violated in the former case,
whereas they are all satisfied in the latter.\footnote{Note that for $N=12$, when non-consecutive fermions interact some of the inequalities are saturated. This happens because all the $I_3$ involving single spin subregions vanish, as well as all the $I_4$ apart from $I_4(2,3,5,6), I_4(1,3,4,6), I_4(1,2,4,5)$, where $i=1,\dots,6$ denotes the i-th Dirac fermion.}

\section{Conclusions}
\label{sec:conclusions}
Based on our numerical results, 
the thermal state $\rho_\beta$ in the sparse SYK model up to {$N \leq 12$} Majorana fermions
satisfies \textit{all} the entropy inequalities involving {$\mathcal{N} \leq 5$} partitions for any temperature $\beta^{-1}$ and sparseness $p$.

We stress that the notion of entanglement entropy we have considered involves different flavors of Majorana fermions, instead of disjoint space regions. While the latter has a clear holographic interpretation in terms of the area of extremal surfaces in the bulk spacetime, the former is not yet understood in this perspective. 
Therefore, it is non-trivial that the same inequalities which are satisfied by the holographic entanglement entropy also hold for the flavor space entanglement entropy.
In the context of QFTs, the holographic entropy inequalities can be regarded as a necessary, but not sufficient, condition for the existence of a gravity dual. Violation of (even one!) inequalities implies that the corresponding QFT state has no smooth gravity dual. On the contrary, if all the inequalities are satisfied we cannot certainly conclude that the QFT state admits a gravity dual.
In light of this, our results are not in contradiction with the literature.
Even assuming that, similar to spatial entanglement, flavor space entanglement plays a role in holography, the fact that all entropy inequalities are satisfied for any number of Majorana fermions $N$, temperature $\beta^{-1}$, and sparseness $p$, does not imply that sparse SYK necessarily has a gravity dual. 

Various other quantities such as the spectral statistics, the spectral form factor, and the OTOC show a marked difference in the case of extremely sparse SYK compared to the full SYK model. In particular for sufficiently sparse SYK (when number of couplings is $\order{(N)}$), the level spacing follows a Poisson distribution as opposed to a Wigner distribution and the ramp structure in the SFF disappears \cite{Anegawa:2023vxq}. Further, the OTOCs no longer saturate the Lyapunov exponent, all of which indicate that such a sparse model will not be holographically dual to a gravitational spacetime. 
Thus, our results for the entropy inequalities neither guarantee nor exclude the existence of a gravity dual, but nevertheless are interesting as the full SYK model satisfies all the inequalities expected for a holographic dual.  

We stress that we have not found any example in which some of the inequalities are satisfied and some are violated. In other words, whenever $I_3 \leq 0$, \textit{all} the higher-party inequalities hold. 
As a consequence, as far as we consider a \textit{typical} entangled state with $I_3 \leq 0$, we expect that all the higher-party inequalities are automatically satisfied.
Since this represents a necessary condition for the existence of a gravity dual, in turn holographic states in QFT should have a typical entangled structure. 
In other words, \textit{atypical states in QFT are not expected to admit a holographic dual}.

As an open question, it would be interesting to investigate the interpretation of the multipartite information $I_n$ in JT gravity. Since in the holographic bulk we have only flavor-singlet observables, black holes are the only objects which contain $N$ flavors. A good analogy is a color index in the large-$N$ QCD. In the holographic bulk, we have only color singlet observables. However, black holes are the only objects which contain $\order{\left(N^2\right)}$ color degrees of freedom in the deconfinement phase. Thus, the flavor space entanglement is a sort of entanglement between black holes. It would be interesting to investigate the role of multipartite entanglement in a black hole viewpoint.  
Also, it would be worth extending our study to the random matrix theory, of which SYK itself is a sparse version, and large-$N$ matrix quantum mechanics instead of vector models. 
Finally, we point out that even though the holographic entropy inequalities cannot distinguish sparse SYK from full SYK, the fact that all the inequalities are satisfied in SYK is an interesting result. This suggests that the entropy inequalities, usually proven using RT formula in field theories with spatial extent, should have a more general proof for holographic theories, which should even be applicable for the case of quantum mechanical systems like the one we studied. An intuitive proof must be along the lines of typicality of the states in the holographic field theory being dual to gravity.

\acknowledgments
The work of NI, SS, NZ is supported by MEXT KAKENHI Grant-in-Aid for Transformative Research Areas A “Extreme Universe” No. 21H05184. The work of NI is also  supported in part by JSPS KAKENHI Grant Number 18K03619. We also acknowledge hospitality at the conference ``Quantum  Information, Quantum  Field Theory and Gravity" (ICTS/qftg2024/08) at ICTS-TIFR, India. SS also acknowledges the support and hospitality at APCTP, Korea. AM thanks Nandini Trivedi for her support during this work. AM also acknowledge the use of the Unity Cluster of the College of Arts and Sciences at the Ohio State University.

\appendix
	
\section{Chirality and degeneracy of the spectrum}
\label{app:degeneracy}

For a system of $n = N/2$ Dirac fermions, it is useful to introduce the following operators:
\begin{itemize}
 \item {\bf Chirality operator.} Defined by
 \[
 \gamma = (2 i)^n \prod_{a=1}^{2n} \psi_a
 = \prod_{a=1}^n ( 2 N_a -1) \, ,
 \]
 the chirality operator has spectrum $\{ -1,1 \}$.
 Given a Hamiltonian of the form $H \sim \left(\psi\right)^q$, we have $[ \gamma, H ]=0$ for any even $q$.
 In the representation in terms of Dirac fermions, the chirality operator `counts' the number of spins down:
 \[
 \gamma \ket{\uparrow \downarrow \downarrow \dots} = (-1)^{\# \, {\rm spin} \, \downarrow} \ket{\uparrow \downarrow \downarrow \dots} \, .
 \]
  \item {\bf Fermion number operator.} Defined by
 \begin{align}
 \label{eq:def-Q}
 Q = \frac{n}{2} + i \sum_{a=1}^n \psi_{2a-1} \psi_{2a}
 = \sum_{a=1}^n N_a
 \end{align}
 The fermion number parity $(-1)^Q$, and not $Q$ itself, is a conserved quantity.
 The fermion number operator `counts' the number of spins up:
 \[
 Q \ket{\uparrow \downarrow \downarrow \dots} = (\# \, {\rm spin} \, \uparrow) \ket{\uparrow \downarrow \downarrow \dots} \, .
 \]
 \item {\bf Particle-hole symmetry operator.} Defined by
 \[
 P = 2^{n/2} K \prod_{a=1}^n \psi_{2a} = K \prod_{a=1}^n (c_a + \bar{c}_a) \, ,
 \]
 with $K$ the anti-linear complex conjugation operator acting as $K(z) = \bar{z}$ for $z \in \mathbb{C}$.
 Again, $[ P, H ]=0$ for any even $q$. Instead, $[ P, \gamma ] =0$ if $n$ is even, whereas $\{ P, \gamma \} =0$ if $n$ is odd.  We have $\{ P, Q \} = n P$. Equivalently, the particle-hole symmetry operator `flips' all spins, i.e. it maps a state with fermion number $Q_0$ to a state with fermion number $n - Q_0$:
 \[
 P \ket{\uparrow \downarrow \downarrow \dots} \propto 
 \ket{\downarrow \uparrow \uparrow \dots} \, .
 \]
 \end{itemize}
 In this appendix, we prove the above-mentioned properties and we exploit them to determine the degeneracy of the spectrum of the full SYK model.

\subsection{Chirality operator}
The chirality operator\footnote{From the explicit form of the Majorana fermions in eqs.~\eqref{eq:psi-odd} and \eqref{eq:psi-even}, it is easy to see that
\begin{align}
\psi_{2j-1} \psi_{2j} = \frac{i}{2} \, \overset{j-1}{\underset{a=1}\otimes} \mathbb{I}^a \otimes \sigma_z^j \overset{n}{\underset{a=j+1}\otimes} \mathbb{I}^a \, , \qquad 1 \leq j \leq n \, .
\end{align}
Therefore, the explicit form of the chirality operator in the chosen basis is
\begin{align}
\gamma = (2i)^n \left( \frac{i}{2} \right)^n \overset{n}{\underset{a=1}\otimes} \sigma_z^a
= i^{2n} \overset{n}{\underset{a=1}\otimes} \sigma_z^a \, .
\end{align}}
\begin{align}
\gamma = (2i)^{n} \prod_{a=1}^{2n} \psi_a \, ,
\end{align}	
satisfies $\gamma^2 = \mathbb{I}$. 
Consequently, the spectrum of the chirality operator $\gamma$ is $\{1, -1 \}$.
From the Clifford algebra \eqref{eq:clifford}, for any $1 < a < 2n$ we get
\begin{align}
\gamma \, \psi_a &= (2i)^n \psi_1 \dots \psi_{2n} \, \psi_a 
= (-1)^{2n-a} (2i)^n \psi_1 \dots \psi_{a-1} \, \psi_a \, \psi_a \, \psi_{a+1} \dots \psi_{2n} \nonumber \\
&= (-1)^{a-1} (-1)^{2n-a} (2i)^n \psi_a \, \psi_1 \dots \psi_{2n}
= (-1)^{2n-1} \psi_a \, \gamma = - \psi_a \, \gamma \, .
\end{align}
The same property trivially holds for $a=1,2n$. In other words, we have $\{ \gamma, \psi_a \} =0$ for any $a=1,\dots, 2n$.
 As a direct consequence,
\begin{align}
\gamma \, \psi_a \psi_b \psi_c \psi_d = (-1)^4 \psi_a \psi_b \psi_c \psi_d \, \gamma = 
\psi_a \psi_b \psi_c \psi_d \, \gamma \, ,
\end{align}
or equivalently $\left[ \gamma, \psi_a \psi_b \psi_c \psi_d \right]=0$, for any $a,b,c,d = 1, \dots, 2n$.
Therefore, we conclude that
\begin{align}
\left[ \gamma, H \right] =0 
\end{align}
for both the (sparse) SYK \eqref{eq:H-syk} and the vector models \eqref{eq:H-vec1},\eqref{eq:H-vec2}.\footnote{Note that this holds for any even $q$. In general, we have $\gamma \, \psi^q = (-1)^q \psi^q \, \gamma$. So, if $q$ is odd, $\{ \gamma, H \} =0$.}
Then, the energy eigenstates can be labeled by their chirality:
\begin{align}
H \ket{E_{j,\alpha}} = E_{j,\alpha} \ket{E_{j,\alpha}} \, , 
\qquad
\gamma \ket{E_{j,\alpha}} = \alpha \ket{E_{j,\alpha}}  \, ,
\qquad 
\alpha=+,- \, .
\end{align}
\\

\subsection{Particle-hole symmetry operator}
Another useful operator is \cite{Fu:2016yrv,Fidkowski:2010jmn,You:2016ldz,Cotler:2016fpe}
\begin{align}
P = 2^{n/2} K \, \prod_{l=1}^n \psi_{2l} \, ,
\end{align}
where $K$ is the anti-linear complex conjugation operator.
We take the same convention as \cite{Cotler:2016fpe}, in which the complex conjugation operator leaves unchanged the Majorana fermions $\psi_{2a}$ and changes the sign of the Majorana fermions $\psi_{2a-1}$, with $a=1, \dots, n$:\footnote{With this choice, the complex Dirac fermions $c_a$ and $\bar{c}_a$ defined in eq.~\eqref{eq:creator-annihilator} are real under conjugation.} 
\begin{align}
\label{eq:K-action}
K \psi_l K = (-1)^l \psi_l \, , 
\qquad
l =1, \dots, 2n \, .
\end{align}
Since $P$ contains a product of $n$ Majorana fermions with even index, we find that 
\begin{align}
P^2 &= 2^n K \, \psi_2 \dots \psi_{2n-2} \psi_{2n} \, K \, \psi_2 \dots \psi_{2n-2} \, \psi_{2n} \nonumber \\
&= 2^n \psi_2 \dots \psi_{2n-2} \psi_{2n} \, \psi_2 \dots \psi_{2n-2} \, \psi_{2n} \nonumber \\
&= 2^n (-1)^{n-1} \psi_2 \dots \psi_{2n-2} \, \psi_2 \dots \psi_{2n-2} \, \psi_{2n}^2 \nonumber \\
&= \dots = 2^k (-1)^{k-1} (-1)^{k-2} \dots (-1) \psi_2^2 \dots \psi_{2k-2}^2 \psi_{2k}^2 \nonumber \\
&= (-1)^{\sum_{j=1}^{n-1} j} \, \mathbb{I} = (-1)^{n(n-1)/2} \, \mathbb{I} \, .
\end{align}
Explicitly, we have
\begin{align}
\label{eq:P-square}
P^2 =
\begin{system}
+\mathbb{I} \qquad \text{if $n$ mod $4 =0$} \\
+\mathbb{I} \qquad \text{if $n$ mod $4 =1$} \\
-\mathbb{I} \qquad \text{if $n$ mod $4 =2$} \\
-\mathbb{I} \qquad \text{if $n$ mod $4 =3$} 
\end{system} \, .
\end{align}
Moreover, note that for $j=1, \dots, n$:
\begin{align}
P \psi_{2a} P &= 2^n K \, \left(\prod_{l=1}^n \psi_{2l} \right) \psi_{2a} \, K \, \prod_{l=1}^n \psi_{2l}
= 2^n \left( \prod_{l=1}^n \psi_{2l} \right) \psi_{2a} \prod_{l=1}^n \psi_{2l}  \nonumber \\
&= (-1)^{a-1}  \, 2^n \left( \prod_{l=1}^n \psi_{2l} \right) \psi_2 \dots \psi_{2a} \, \psi_{2a} \, \psi_{2a+2} \dots \psi_{2n} \nonumber \\
&= (-1)^{a-1} (-1)^{n-a} \, 2^n \left( \prod_{l=1}^n \psi_{2l} \right) \left( \prod_{l=1}^n \psi_{2l} \right) \psi_{2a} \nonumber \\
&= (-1)^{n-1} P^2 \psi_{2a} \, ,
\end{align}
\begin{align}
P \psi_{2a-1} P &= 2^n K \, \left(\prod_{l=1}^n \psi_{2l} \right) \psi_{2a-1} \, K \, \prod_{l=1}^n \psi_{2l}
= - 2^n \left( \prod_{l=1}^n \psi_{2l} \right) \psi_{2a-1} \prod_{l=1}^n \psi_{2l}  \nonumber \\
&= - (-1)^n \, 2^n \left( \prod_{l=1}^n \psi_{2l} \right) \left( \prod_{l=1}^n \psi_{2l} \right) \psi_{2a-1} \nonumber \\
&= (-1)^{n+1} P^2 \psi_{2a-1} \, ,
\end{align}
where use has been made of eq.~\eqref{eq:K-action}.
Then, we have
\begin{align}
P \psi_l P = (-1)^{n-1} P^2 \psi_l \, ,
\qquad
l=1, \dots, 2n \, .
\end{align}
By applying $P$ to both sides,
this can be written as
\begin{align}
\label{eq:P-action}
P \psi_l P^2 =  (-1)^{n-1} P^2 \psi_l P
\qquad \implies \qquad
P \psi_l = (-1)^{n-1} \psi_l P \, ,
\end{align}
since $P^2 \propto \mathbb{I}$.
It is now easy to see that
\begin{align}
P \psi_a \psi_b \psi_c \psi_d = (-1)^{4(n-1)} \psi_a \psi_b \psi_c \psi_d P = \psi_a \psi_b \psi_c \psi_d P
\end{align}
for any $a,b,c,d = 1, \dots, 2n$, which implies that
\begin{align}
\left[ P, H \right] = 0 \, .
\end{align}
Therefore, $P$ leaves the energy of a state invariant.
\\

\subsection{Degeneracy of the spectrum}
By eq.~\eqref{eq:P-action}, we find
\begin{align}
P \, \gamma &= P (2i)^n \prod_{l=1}^{2n} \psi_l
= (-2i)^n P \prod_{l=1}^{2n} \psi_l \nonumber \\
&= (-1)^{2n(n-1)} (-2i)^n \left( \prod_{l=1}^{2n} \psi_l \right) P
= (-1)^n \gamma \, P \, ,
\end{align}
where in the first line we have used the anti-linearity of $K$. 
Therefore, if $n$ is even $\left[ P, \gamma \right] =0$, whereas if $n$ is odd $\{ P, \gamma \} =0$.
Let us now consider an energy eigenstate $\ket{E_{j,\alpha}}$ with chirality $\alpha =+,-$. 
We distinguish among different cases:
\begin{itemize}
 \item {\bf $n=N/2$ is odd ($N=6, 10, 14, \dots$).}  
 Since $P$ and $\gamma$ anti-commute,
 \begin{align}
 \gamma (P \ket{E_{j,\alpha}}) = -P \gamma \ket{E_{j,\alpha}} 
 = - \alpha P \ket{E_{j,\alpha}} \, .
 \end{align}
 In other words, $P$ changes the chirality of the energy eigenstates, but leaves the energy invariant. 
 Then, for every energy eigenstate there exists another eigenstate with the same energy and opposite chirality.
 In particular, there are two degenerate ground states with opposite chirality.
 \item {\bf $n =N/2$ is even and $n$ mod $4=2$ ($N=4, 12, 20, \dots$).}
 Since $P$ and $\gamma$ commute, 
 \begin{align}
 \gamma (P \ket{E_{j,\alpha}}) = P \gamma \ket{E_{j,\alpha}} 
 = \alpha P \ket{E_{j,\alpha}} \, .
 \end{align}
 So, $P$ leaves both the chirality and the energy of the Hamiltonian eigenstates invariant.
 Now, suppose that $P$ maps an eigenstate into itself: $P \ket{E_{j,\alpha}} = q \ket{E_{j,\alpha}}$.
 Then, by the anti-linearity of $P$,
 \begin{align}
 P^2 \ket{E_{j,\alpha}} = P ( P \ket{E_{j,\alpha}} ) = P (q  \ket{E_{j,\alpha}} ) = \bar{q} \, P \ket{E_{j,\alpha}} = |q|^2 \ket{E_{j,\alpha}} \, ,
 \end{align}
 which is in contradiction with $P^2 = -\mathbb{I}$.
 Consequently, $P$ necessarily maps an eigenstate of the Hamiltonian to a different eigenstate with the same energy  and chirality. There is thus a two-fold degeneracy in every chirality class. In particular, there are two degenerate ground states with the same chirality. 
 \item {\bf $n=N/2$ is even and $n$ mod $4=0$ ($N=8, 16, 24, \dots$).}
 Contrary to the previous case, $P^2 = \mathbb{I}$. So, $P$ can map an egeinstate into itself, and degeneracy is not guaranteed.
\end{itemize}
See Fig.~\ref{fig:degeneracy} for a schematic representation of the degeneracy and chirality of the energy eigenstates.

\section{Tripartite information for large temperature}
\label{app:I3-large-T}

We here determine an expression for the tripartite information of the thermal state at small $\beta$.
If we take $\rho$ to be the thermal state $\rho_\beta$ in eq.~\eqref{eq:thermal-state} on the full system $ABC$, then the matrices appearing in eq.~\eqref{eq:rho-small-beta} are simply $M_1=H$ and $M_2 = H^2$, so that
\begin{align}
S_{ABC} &= \log d_{ABC}+ \frac{(\tr (H))^2 - d_{ABC} \tr (H^2)}{2 d_{ABC}^{2}} \beta^2 + \mathcal{O}(\beta^3) \\
&= n \log 2 - \frac{\sum_{i <j} (E_i - E_j)^2}{2^{2n+1}} \beta^2 + \mathcal{O}(\beta^3)  \nonumber \, ,
\end{align}
where in the second line we have used the result in eq.~\eqref{eq:S-small-beta}.
The same expression for $S_{ABC}$ can be obtained by expanding the thermal entropy in eq.~\eqref{eq:thermal-entropy}.
\\
Let us now evaluate the reduced density matrix
\begin{align}
\rho_{AB} &= \tr_C (\rho_\beta) \sim   \frac{\tr_C \left( \mathbb{I} - \beta H + \frac{\beta^2}{2} H^2  \right) }{\tr \left( \mathbb{I} - \beta H + \frac{\beta^2}{2} H^2 \right)} 
=  \frac{d_c \, \mathbb{I} - \beta \tr_C (H) + \frac{\beta^2}{2} \tr_C (H^2) }{\tr \left( \mathbb{I} - \beta H + \frac{\beta^2}{2} H^2 \right)} \\
&= \frac{d_c \, \mathbb{I} - \beta \tr_C (H) + \frac{\beta^2}{2} \tr_C (H^2) }{\tr \left( d_c \, \mathbb{I} - \beta \tr_C (H) + \frac{\beta^2}{2} \tr_C (H^2) \right)} \\
&= \frac{\mathbb{I} - \beta \tr_C (H)/d_c + \frac{\beta^2}{2} \tr_C (H^2)/d_c }{\tr \left( \mathbb{I} - \beta \tr_C (H)/d_c + \frac{\beta^2}{2} \tr_C (H^2)/d_c \right)}  \, .
\end{align}
Here $\tr_C$ denotes the partial trace over the subsystem $C$, whereas $\tr$ denotes the full trace.
In the second line, we have used the known fact that the partial trace preserves the full trace:
$\tr (M) = \tr (\tr_C (M))$.
From the last line, it is clear that $\rho_{AB}$ has the same form as eq.~\eqref{eq:rho-small-beta},
with $M_1 = \tr_C (H)/d_C$.
Therefore, eq.~\eqref{eq:S-small-beta} still applies and gives
\begin{align}
S_{AB} &= \log d_{AB} + \frac{(\tr ( \tr_C (H)/d_C))^2 - d_{AB} \tr ( (\tr_C (H)/d_C)^2)}{2 d_{AB}^2} \beta^2 + \mathcal{O}(\beta^3) \\
&= \log d_{AB} + \frac{(\tr (H))^2 - d_{AB} \tr ( (\tr_C (H))^2)}{2 d_{ABC}^2} \beta^2 + \mathcal{O}(\beta^3) \, ,
\end{align}
where in the second equality we have used the property $d_{ABC} = d_{AB} \, d_C$, which holds for any bipartition.
Following the same argument, $\rho_A = \tr_{BC} (\rho_\beta)$ is also of the form in eq.~\eqref{eq:rho-small-beta} with $M_1 = \tr_{BC}(H)/d_{BC}$, so that
\begin{align}
S_A = \log d_{A} + \frac{(\tr (H))^2 - d_{A} \tr ( (\tr_{BC} (H))^2)}{2 d_{ABC}^2} \beta^2 + \mathcal{O}(\beta^3) \, .
\end{align}
Note that, in general, $\tr_C (H)$ and $\tr_{BC} (H)$ are not diagonal in the basis of eigenstates of $H$.
Putting all together, we find eq.~\eqref{eq:I3-small-beta}.

\section{Negativity of the tripartite information for random states}
\label{negtrp}
In this Appendix, we shall analyze a simple scenario to show that when the ground states are a typical linear combination of ``partition basis" states and the density matrix is of the form of eq.~\eqref{rinfv2}, then the tripartite information is negative. 
We consider the case of $N=6$ for the analysis performed here. This corresponds to three spins, which constitute the three parties. 
Let us consider two orthonormal random states as below:
\begin{align}
|	A\rangle =\frac{c_1 \ket{\uparrow\uparrow\uparrow}+c_4\ket{\downarrow\downarrow\uparrow}+c_6\ket{\downarrow\uparrow\downarrow}+c_7\ket{\uparrow\downarrow\downarrow}}{\sqrt{\abs{c_1}^2+\abs{c_4}^2+\abs{c_6}^2+\abs{c_7}^2}} \, , \nonumber\\
|   B\rangle=\frac{c_2 \ket{\downarrow\uparrow\uparrow}+c_3\ket{\uparrow\downarrow\uparrow}+c_5\ket{\uparrow\uparrow\downarrow}+c_8\ket{\downarrow\downarrow\downarrow}}{\sqrt{\abs{c_2}^2+\abs{c_3}^2+\abs{c_5}^2+\abs{c_8}^2}} \, , \label{gs1}
\end{align}
where $c_1,\dots c_8$ are random complex numbers drawn from a uniform distribution with real and imaginary parts in the range $[0,1]$. Note that the two states $\ket{A}$ and $\ket{B}$ have fixed and opposite chirality. Therefore, we can consistently regard them as possible ground states of the sparse SYK model with $N=6$.
We then take the following density matrix: 
\begin{align}
	\rho=\half \left( \ketbra{A} +\ketbra{B} \right) \, .
\end{align}
A computation of $I_3$ in this case for various random values of $c_1\dots c_8$ shows that the tripartite information is almost always negative. In particular, we have examined an ensemble size of 1000 and we have found that for $\gtrapprox 990$ members of the ensemble, $I_3<0$. This is in line with some comments made at the end of section \ref{sec:tripartite-info}, where we argued that when the ground states are complicated linear combinations of ``partition" basis states, the tripartite information would be negative.

\section{Independent entropy inequalities for five partitions}
\label{app:inequalities}

In this Appendix, we provide details about the independent entropy inequalities for $\mathcal{N} = 5$ parties and their check in the SYK model. For simplicity, we focus on $N=12$ Majorana fermions, or equivalently $n=6$ spins. 
This choice leaves us with two possible partitions of the spins into five groups $A,B,C,D,E$:
\begin{itemize}
 \item Each group consist of a single spin, so that one spin is left out of the partition. 
 \item One of the five groups contains two spins and the remaining four groups consist of a single spin each, so that all spins are involved in the partition.
\end{itemize}
Note that in the second case the system $ABCDE$ is in the thermal state $\rho_{\beta}$, whereas in the first case $ABCDE$ is in a reduction of the thermal state obtained by tracing out the leftover spin. 

\begin{figure}[h]
\centering
\includegraphics[scale=0.25]{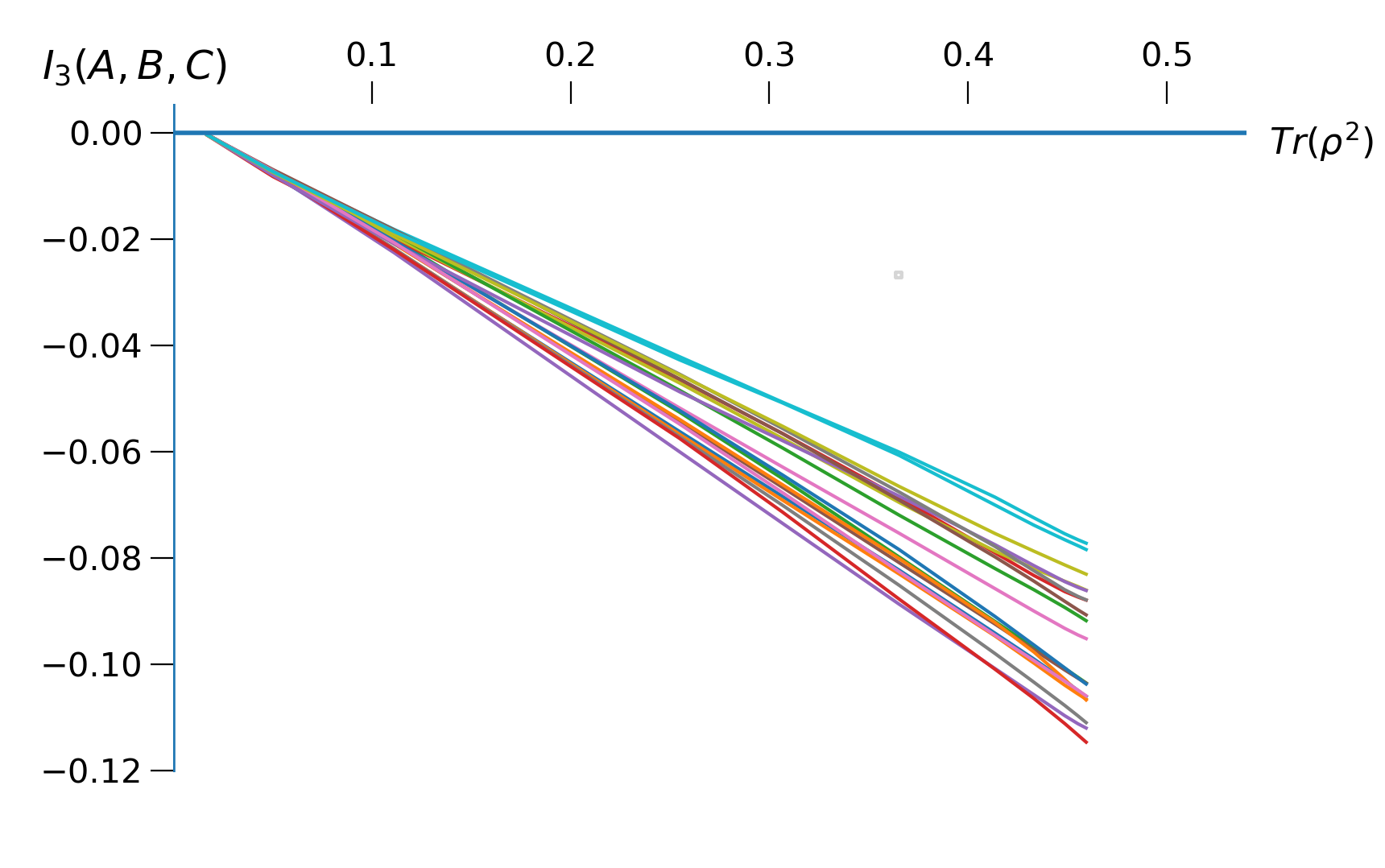} 
\qquad
\includegraphics[scale=0.6]{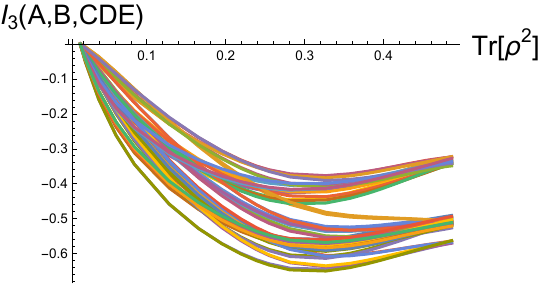} 
\caption{Tripartite information $I_3(A,B,C)$ and $I_3(A,B,CDE)$ in the full SYK for $A,B,C,D,E$ containing one spin each.}
\label{fig:second-class-single}
\end{figure}

\begin{figure}[h]
\centering
\includegraphics[scale=0.6]{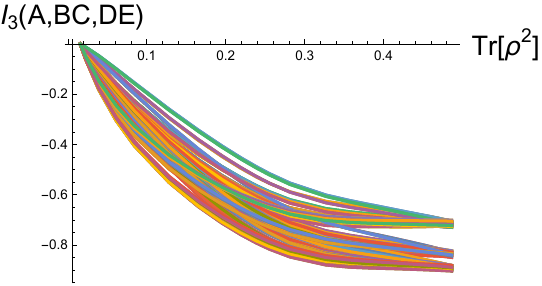} 
\qquad
\includegraphics[scale=0.6]{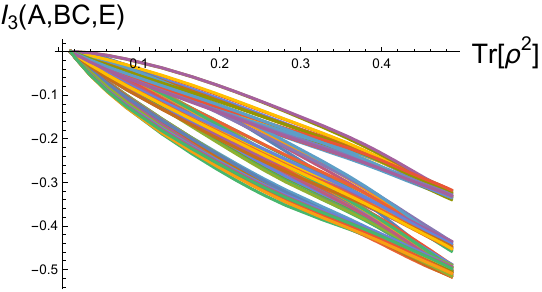} 
\caption{Tripartite information $I_3(A,BC,DE)$ and $I_3(A,BC,E)$ in the full SYK for $A,B,C,D,E$ containing one spin each.}
\label{fig:third-class-single}
\end{figure}

\begin{figure}[h]
\centering
\includegraphics[scale=0.48]{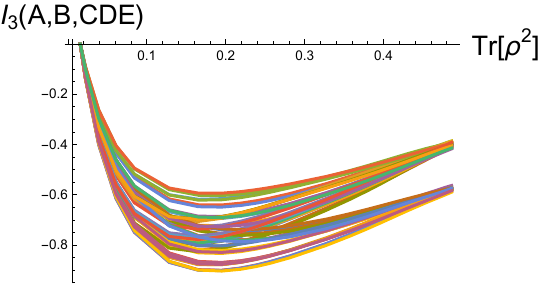} 
\qquad
\includegraphics[scale=0.48]{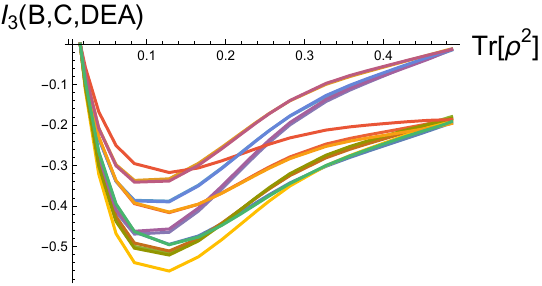} 
\qquad
\includegraphics[scale=0.48]{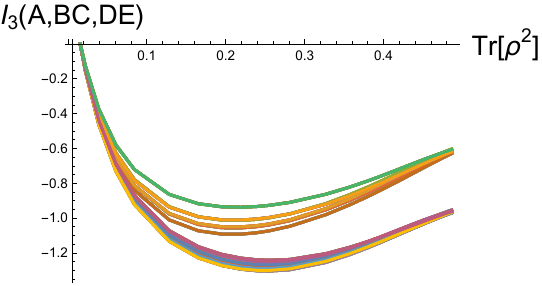} 
\caption{Tripartite information $I_3(A,B,CDE)$, $I_3(B,C,DEA)$, and $I_3(A,BC,DE)$ in the full SYK for $A$ containing two spins and $B,C,D,E$ one spin each, including all the permutations.}
\label{fig:second-third-class-double}
\end{figure}

\begin{figure}[h]
\centering
\includegraphics[scale=0.55]{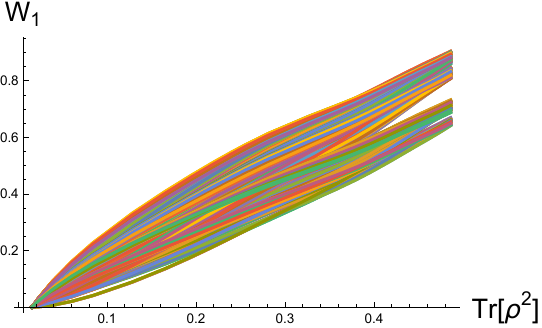} 
\qquad
\includegraphics[scale=0.55]{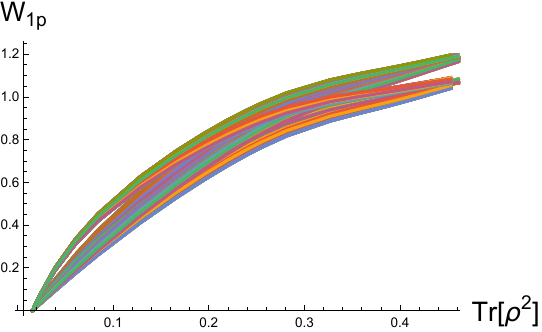} 
\qquad
\includegraphics[scale=0.55]{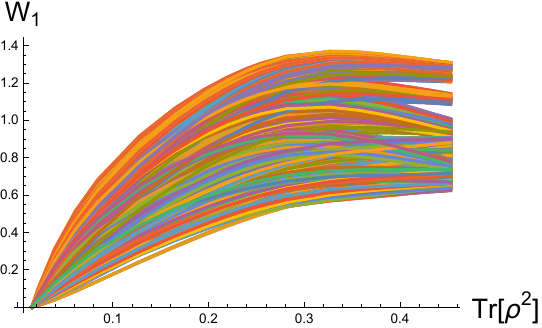} 
\qquad
\includegraphics[scale=0.55]{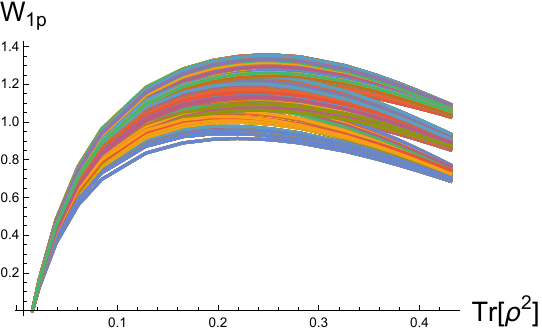} 
\caption{Quantities $W_1$ and $W_{1p}$ in the full SYK for $A,B,C,D,E$ containing one spin each (top), and for $A$ (and permutations) containing two spins (bottom).}
\label{fig:fourth-class}
\end{figure}

\begin{figure}[h]
\centering
\includegraphics[scale=0.55]{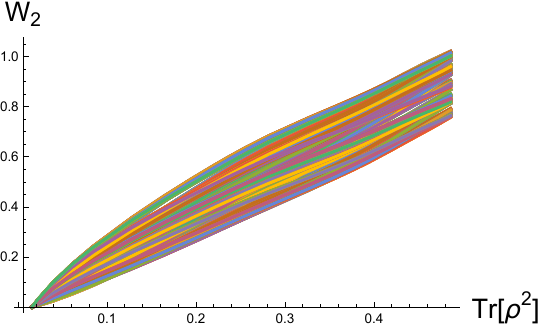} 
\qquad
\includegraphics[scale=0.55]{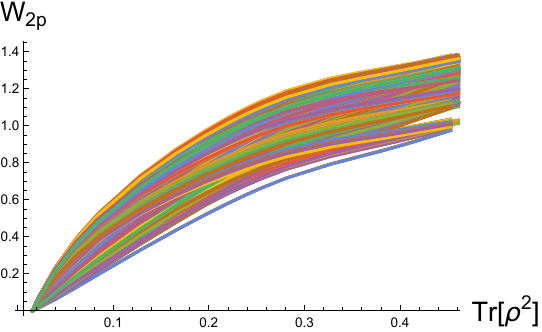} 
\qquad
\includegraphics[scale=0.55]{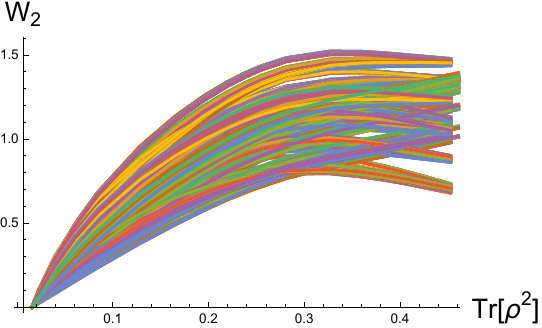} 
\qquad
\includegraphics[scale=0.55]{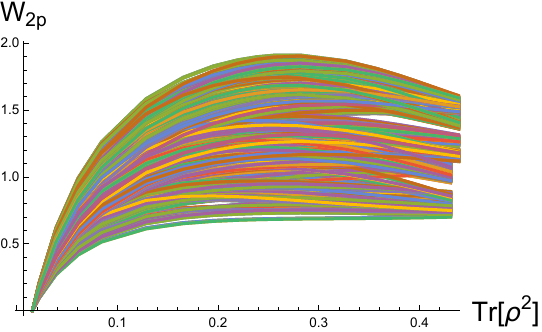} 
\caption{Quantities $W_2$ and $W_{2p}$ in the full SYK for $A,B,C,D,E$ containing one spin each (top), and for $A$ (and permutations) containing two spins (bottom).}
\label{fig:fifth-class}
\end{figure}

\begin{figure}[h]
\centering
\includegraphics[scale=0.55]{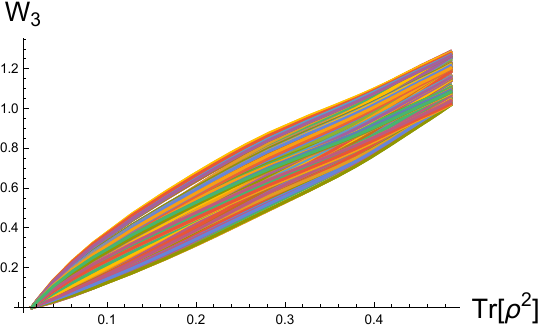} 
\qquad
\includegraphics[scale=0.55]{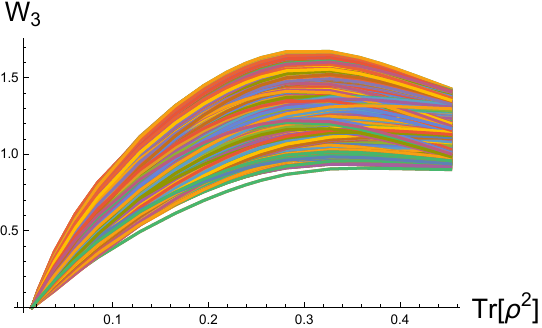} 
\qquad
\includegraphics[scale=0.55]{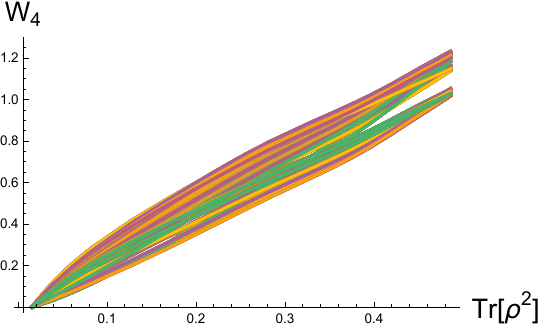} 
\qquad
\includegraphics[scale=0.55]{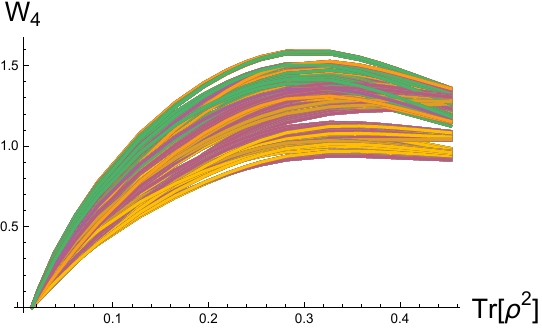} 
\qquad
\includegraphics[scale=0.55]{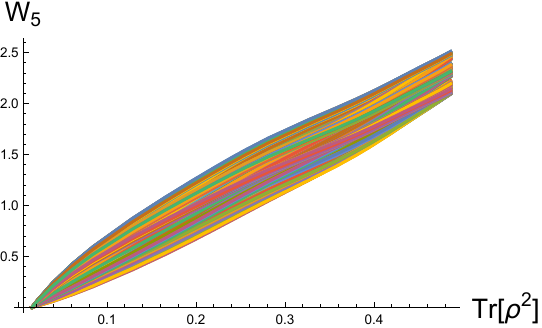} 
\qquad
\includegraphics[scale=0.55]{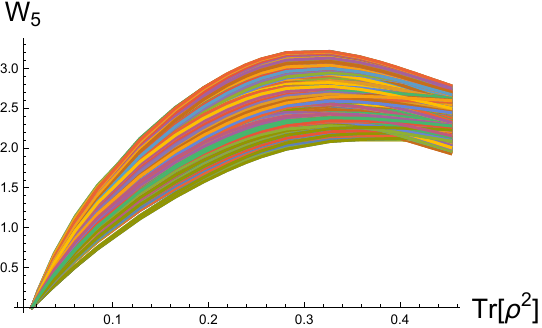} 
\caption{Quantities $W_3, W_4,$ and $W_5$ in the full SYK for $A,B,C,D,E$ containing one spin each (left), and for $A$ (and permutations) containing two spins (right).}
\label{fig:last-classes}
\end{figure}

Let us now discuss the eight classes of entropy inequalities for $\mathcal{N}=5$ parties $A,B,C,D,E$.
In the following discussion, we assume that the system $ABCDEP$, with $P$ the purifier, is indeed in a pure state.
\\
The first class of inequalities corresponds to the subadditivity of the entanglement entropy $I_2(A,B) \geq 0$,  for any pair of subregions chosen from $A,B,C,D,E$. In addition, we must include the purifier $I_2(A,P) \geq 0$, or equivalently $S_{ABCDE} + S_A - S_{BCDE} \geq 0$, obtaining all the independent Araki-Lieb inequalities. Since all the inequalities in the first class are known to be true in any quantum system, we do not show their behavior in the SYK model.
\\
The second class of inequalities contains the monogamy of mutual information $I_3(A,B,C) \leq 0$ involving any possible group of three partitions, and the purifier together with a pair of partitions $I_3(A,B,P) \leq 0$, which is equivalent to $I_3(A,B,CDE) \leq 0$.
In Fig.~\ref{fig:second-class-single}, we display such inequalities for $A,B,C,D,E$ consisting of one spin each.
\\
The third class of inequalities corresponds to the uplift of the monogamy of mutual information $I_3(A,BC,DE) \leq 0$, together with the cyclic permutations of the five subregions. Note that $I_3(P,BC,DE) = I_3(A,BC,DE)$, so the corresponding inequality is already included in the previous subset. The purifier plays instead a role in inequalities of the form $I_3(A,BC,DP) \leq 0$, which is equivalent to $I_3(A,BC,E) \leq 0$.
In Fig.~\ref{fig:third-class-single}, we show all the independent inequalities in the third class for $A,B,C,D,E$ consisting of a single spin each.
\\
In Fig.~\ref{fig:second-third-class-double} we plot all the quantities in the second and third class of inequalities when one of the five subregions involved contains two spins.
\\
We now turn to the new classes of inequalities appearing for five partitions.
The fourth class contains all the permutations of 
\begin{align}
\label{eq:W1-app}
W_1 &= I_4 (A,B,C,D) + I_4 (A,C,D,E) - I_3 (A,C,D) - I_3 (A,C,E) - I_3 (B,C,D)   \nonumber \\
&= - I_3(AB,C,D) - I_3(A,C,DE) + I_3(A,C,D)
\geq 0
\, ,
\end{align}
where the second line is obtained by the relation 
$I_4(A,B,C,D) - I_3(A,B,C) - I_3(A,B,D) = - I_3 (A,B,CD)$ \cite{Hernandez-Cuenca:2023iqh}.
It can be checked that including the purifier gives a new subset of inequalities only when we take $C \rightarrow P$ in eq.~\eqref{eq:W1-app}. This leads to
\begin{align}
W_{1p} = - I_3(AB,CE,D) - I_3(A,BC,DE) + I_3(A,D,BCE) \geq 0 \, .
\end{align}
We plot these quantities in Fig.~\ref{fig:fourth-class}.
\\
The fifth class of inequalities includes the permutations of
\small
\begin{align}
\label{eq:W2-app}
W_2 &= I_4(A,B,C,D) + I_4(B,C,D,E) - I_3(A,B,E) - I_3(A,C,D) - I_3(B,C,D) - I_3(C,D,E) \nonumber  \\
&= -I_3(A,C,BD) + I_3(A,B,C) - I_3(C,D,BE) - I_3(A,B,E)
\geq 0 \, .
\end{align}
\normalsize
The substitution $B \rightarrow P$ in eq.~\eqref{eq:W2-app} is equivalent to a proper permutation of $A,B,C,D,E$. On the other hand, any of the substitutions $A,C,D,E \rightarrow P$ leads to an independent inequality. For instance, taking $A \rightarrow P$ we get
\begin{align}
W_{2p} = -I_3(AE,C,BD) + I_3(B,C,ADE) - I_3(C,D,BE) - I_3(ACD,B,E) \geq 0 \, .
\end{align}  
In Fig.~\ref{fig:fifth-class}, we show the quantities belonging to the fifth class of inequalities.
\\
The remaining three classes contain all the instances of
\begin{align}
W_3 = I_{ABCD} + I_{ACDE} + I_{BCDE} - I_{ACD} - I_{ACE} - I_{BCD} - I_{BDE} \geq 0 \, ,
\end{align}
\begin{align}
W_4 = I_{ABCE} + I_{ACDE} + I_{BCDE} - I_{ABD} - I_{ACE} - I_{BCE} - I_{CDE} \geq 0 \, ,
\end{align}
\begin{align}
W_5 &= 2 I_{ABCD} + I_{ABCE} + 2 I_{ABDE} + I_{ACDE}  \nonumber \\
 &  \qquad -I_{ABD} - 2 I_{ABE} - 2 I_{ACD} - I_{ACE} - I_{BCD} - I_{BDE} \geq 0 \, ,
\end{align} 
respectively.
Here we have defined without risk of confusion $I_{ABCD} \equiv I_4(A,B,C,D)$ and $I_{ABC} \equiv I_3(A,B,C)$.
In all these quantities the purifier plays no role, in the sense that any of the substitutions $A,B,C,D,E \rightarrow P$ is equivalent to a proper permutations of the five subregions $A,B,C,D,E$.
We display the quantities in the last three classes in Fig.~\ref{fig:last-classes}.
As a final remark, note that in this appendix we have shown plots for the full SYK ($p=1$) only.
As commented in section \ref{subsec:tripartite-info-syk}, around Fig.~\ref{fig:I3-syk-purity}, with the increasing of sparseness the maximum value of purity gets smaller. Apart from this, the qualitative behavior of the plots remains unchanged. Such a tendency, which we show explicitly in section \ref{subsec:multipartite-info-syk} for one representative of each of the last five classes of inequalities, implies that all the entropy inequalities for $\mathcal{N} =5$ are satisfied for every temperature $\beta^{-1}$ and sparseness $p$ in the sparse SYK model.

\bibliographystyle{JHEP}

\bibliography{bibliography}	
	
\end{document}